\documentclass[american]{article}
\usepackage[T1]{fontenc}
\usepackage[latin9]{inputenc}
\usepackage{geometry}
\usepackage{soul}
\usepackage{array}
\usepackage{rotfloat}
\usepackage{graphicx}
\usepackage{subscript}
\usepackage[authoryear]{natbib}
\usepackage{color}

\usepackage{placeins}
\usepackage{refcount}
\usepackage{pdflscape}
\usepackage[switch]{lineno}
\usepackage{afterpage}
\setcitestyle{aysep={}}

\newcounter{ctSpectrum}
\newcommand{\newspec}[1]{{\refstepcounter{ctSpectrum}\label{#1}\ifnum\value{ctSpectrum}<10 0\fi\arabic{ctSpectrum}}}
\newcommand{\refspec}[1]{\ifnum\getrefnumber{#1}<10 0\fi\ref{#1}}

\newcommand{\ImagesDir}{Images}

\newcommand{\ImageWidthFactor}{0.45}
\newcommand{\PageWidthFactor}{0.95}

\makeatletter

\usepackage{authblk}
\usepackage{sectsty} 
\usepackage{multicol}
\usepackage{microtype}
\usepackage[labelformat=empty]{caption}
\usepackage{makecell}

\makeatother

\usepackage{babel}
\begin{document}

\title{Fine-grained Material Associated with a Large Sulfide  returned from Comet 81P/Wild 2}

\author{}

{\centering
 \maketitle
Z. Gainsforth$^{1,\dagger}$, 
A. J. Westphal$^1$, 
A. L. Butterworth$^1$, 
C. E. Jilly-Rehak$^1$, 
D. E. Brownlee$^2$, 
D. Joswiak$^2$, 
R. C. Ogliore$^3$,
M. E. Zolensky$^4$,
H. A. Bechtel$^5$,
D. S. Ebel$^6$,
G. R. Huss$^7$,
S. A. Sandford$^8$,
A. J. White$^9$    \\\ \\ \normalsize$^1$Space Sciences Laboratory, University of California, Berkeley, CA 94720,
$^2$Dept. of Astronomy, University of Washington, Seattle, WA 98195, 
$^3$Department of Physics, Washington University in St. Louis, St. Louis, MO, 63117,
%
$^4$ARES, NASA Johnson Space Center, Houston, TX 77058,
$^5$Advanced Light Source, Lawrence Berkeley Laboratory, Berkeley, CA 94720,
$^6$Dept. Earth Planet. Sci., American Museum Natural History, NY, NY 10024,
$^7$University of Hawai`i at Manoa, Honolulu, HI 96822,
$^8$NASA Ames Research Center, Moffett Field, CA 94035,
$^9$Dept. Astro. and Planet. Sci., University of Colorado, Boulder, CO 80309
$^{\dagger}$e-mail: zackg@ssl.berkeley.edu \\\ \\
}

\newpage{}
\begin{abstract}
In a consortium analysis of a large particle captured from the coma of comet 81P/Wild 2 by the Stardust spacecraft,  we report the discovery of a field of fine-grained material (FGM) in contact with a large sulfide particle.  The FGM was partially located in an embayment in the sulfide, so appears to have been largely protected from damage during hypervelocity capture in aerogel.  Some of the FGM particles are indistinguishable in their characteristics from common components of chondritic-porous interplanetary dust particles (CP-IDPs), including glass with embedded metals and sulfides (GEMS) and equilibrated aggregates (EAs).   The sulfide exhibits surprising Ni-rich lamellae, which may indicate that this particle experienced a long-duration heating event after its formation but before incorporation into Wild 2.   We discuss the relationship of the FGM to the sulfide, to other Wild 2 particles and to the history of the Solar nebula.
\end{abstract}
\twocolumn
\sectionfont{\centering}

\section*{INTRODUCTION}

Anhydrous, fine-grained  (``chondritic-porous'') interplanetary dust particles (CP-IDPs) are regularly captured by aircraft-borne collectors in the stratosphere.  Several lines of evidence point to a cometary origin for a subset of CP-IDPs \citep{Love:1991uo, 1993LPI....24..205B, Joswiak:2007ve}, and while there is substantial evidence that some specific CP-IDPs are cometary \citep{Joswiak:2017dk}, there is no hard link between specific CP-IDPs and specific comets.   CP-IDPs commonly contain 100-500\,nm amorphous silicates known as GEMS (Glass with Embedded Metals and Sulfides), and crystalline silicate aggregates of similar size and larger called Equilibrated Aggregates (EAs).   The relationship, if any, between GEMS and EAs is unclear, but a subset of EAs have compositions similar to  GEMS and it has been suggested that many EAs are GEMS that have been heated and annealed \citep{Brownlee:2005wh, Keller:2009wa}.  In addition to this fine-grained material (FGM), CP-IDPs also contain larger crystalline silicates, sulfides and, more rarely, metal.  

The Stardust mission returned $\sim$300$\,\mu$g of material from the coma of Jupiter-family comet Wild 2 to terrestrial laboratories, in a collector composed of aerogel and aluminum foil \citep{Brownlee:2006kw, Brownlee:2013gv,Horz:2006p24}.
There was an expectation that the Stardust collection would include abundant GEMS, EAs, and  pre-solar grains. 
The collection was found to contain abundant crystalline particles $\gg$ 1\,$\mu$m in size \citep{Brownlee:2006kw}, but evidence for GEMS and EAs has been elusive, probably because of the violence of  the hypervelocity capture process and the fragility of FGM.  Pre-solar grains have been reported, but the abundance is difficult to estimate accurately because of the uncertainties in survival efficiency \citep{Floss:2013hd}.

Nevertheless, several particles extracted from the aerogel collectors have shown evidence for FGM adhering to their peripheries.  
These include Iris (C2052,12,74), a $\approx$15\,$\mu$m chondrule-like object with associated primitive sulfides and an enstatite whisker \citep{Stodolna:2013co}, and Cecil (C2062,2,162), a particle containing a chondrule-like object and fine-grained sulfides, spinels, and pyroxenes.
In Cecil, the fine grained material was largely embedded in glass suggesting that some heating alteration may have occurred during aerogel capture, but several of the fine grained components retained primitive signatures despite the apparent heating \citep{Gainsforth:2014wc, Gainsforth2015:spinels}.  
\cite{Joswiak2012:thetome} studied a sulfide particle,  Febo (C2009,2,57), with FGM adhering to one side.  
This FGM showed strong evidence for melting, probably caused by aerogel capture.  
However, $^{15}$N-rich organics survived in select locations within the fine-grained material implying that at least some of the material in the Febo FGM retains chemical signatures from before aerogel capture \citep{Matrajt:2008abc}.

Here we report the discovery and analysis of a particle named Andromeda (C2086,22,191) returned from comet 81P/Wild 2 by the Stardust spacecraft.  
It consists of a field of fine-grained material (FGM) in association with a large sulfide particle.  
Some FGM appears to have been protected from significant damage during hypervelocity capture in aerogel and has strong affinities with components of CP-IDPs, including GEMS and EAs.

\section*{METHODS}

\begin{figure*}[tbp!]
  \centering
  \includegraphics[width=\PageWidthFactor\textwidth]{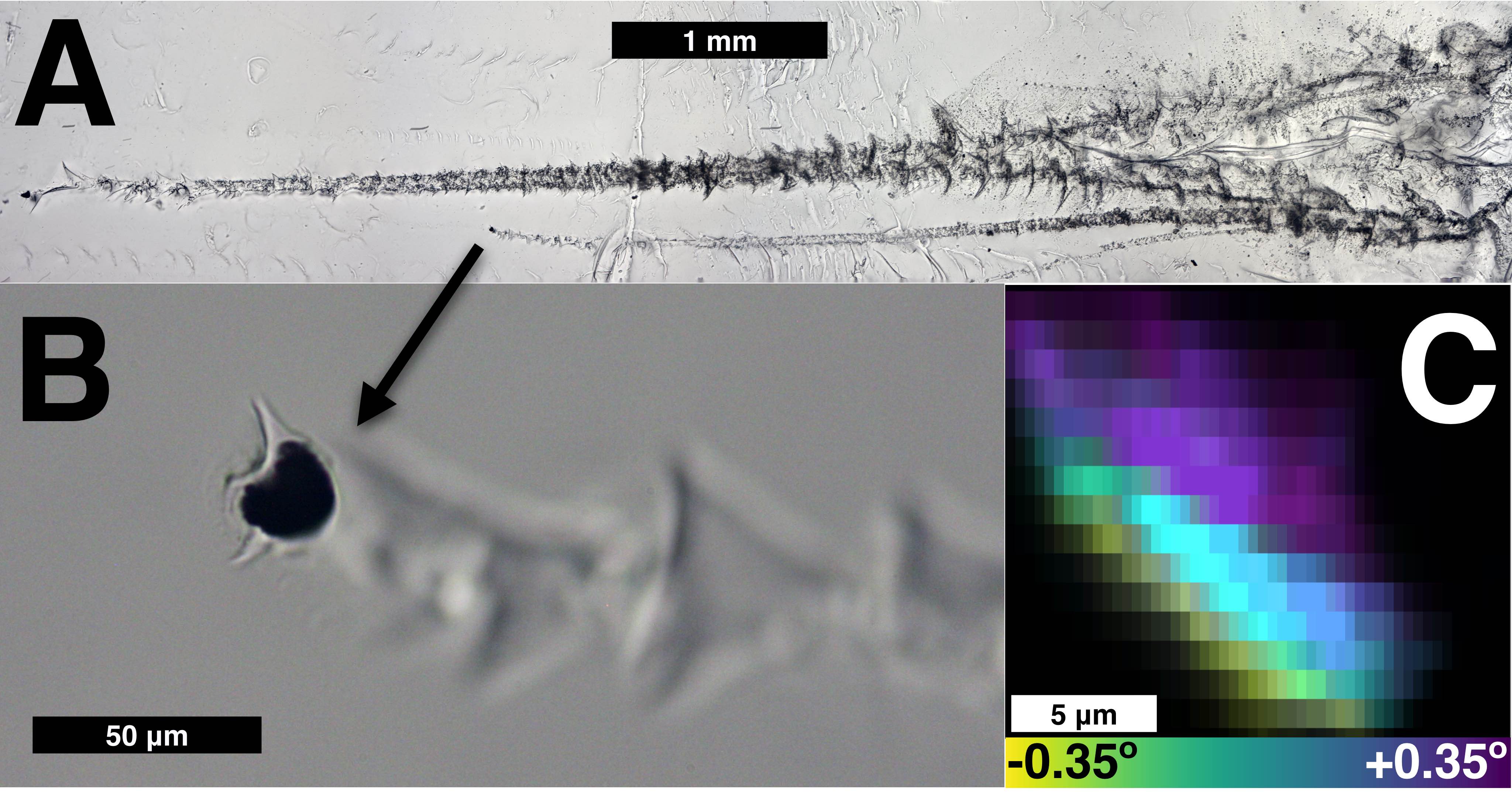}
  \caption{Figure \ref{fig:TrackOptical}.  A) Optical mosaic image of track 191 produced from image stack at varying focus.  Andromeda was found at the end of an 8.8\,mm track.  The arrow connects the location of Andromeda in the track to panel B.  B) Optical zoom-in of Andromeda.  C) XRD topograph of Andromeda showing that the sulfide is a single crystal but with three subgrains spaced in 0.35$^{\circ}$ increments.  Due to sample preparation and beamline geometry panel C is not at the same viewing angle as panels A and B.}
  \label{fig:TrackOptical}
\end{figure*}

\begin{figure*}[thbh!]
  \centering
  \includegraphics[width=\PageWidthFactor\textwidth]{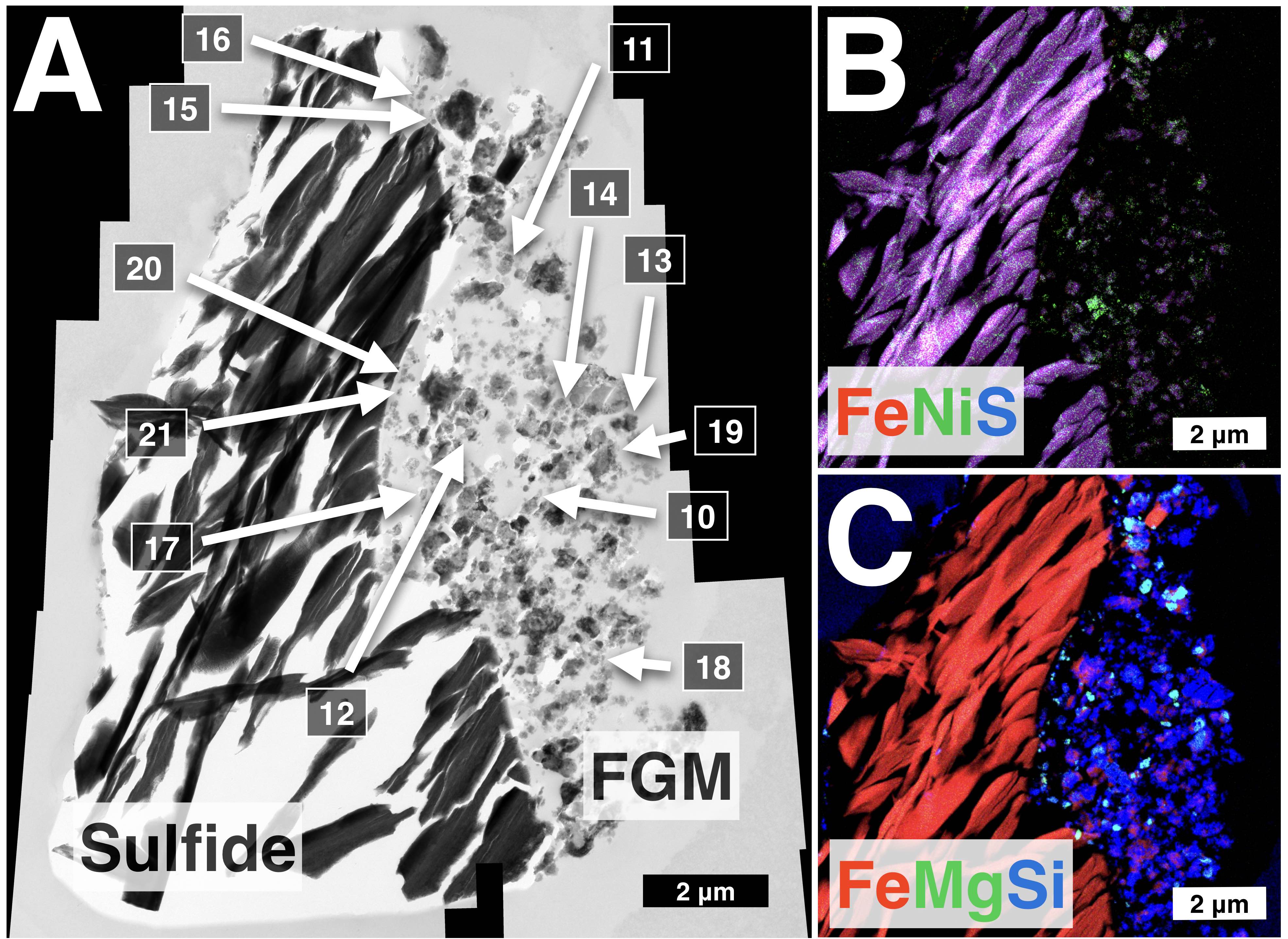}
  \caption{Figure \ref{fig:AndromedaPanorama}.  A) Mosaic of brightfield TEM images of Andromeda showing the sulfide on the left, and fine-grained material (FGM) on the right.  Objects in the FGM are labelled according to their names in Table \ref{tab:NanophaseCompositions}.   B) EDS map of the FGM with iron (red), nickel (green), and sulfur (blue).  The Ni content in the FGM is significantly higher than the Ni content of the primary sulfide impactor.  C) EDS map of the FGM with iron (red), magnesium (green), and silicon (blue). Mg-rich grains include equilibrated aggregates, GEMS and altered objects.  The interstitial space is filled with epoxy.}
  \label{fig:AndromedaPanorama}
\end{figure*}

Andromeda is one of at least seven terminal particles of type-B \citep{Burchell:2008p13442} track 191 (C2086,22,191, Fig. \ref{fig:TrackOptical}A and B).
It was located at a depth of 8.8 mm from the space-exposed surface of the aerogel collector, and consisted of a 15 x 15 x 20\,$\mu$m iron sulfide  with abundant FGM aggregated in an embayment of the sulfide (Fig. \ref{fig:AndromedaPanorama}).
The particle was extracted in an aerogel ``keystone'' \citep{Westphal:2004p13506}, then embedded in EMBED 812 epoxy on the end of an epoxy bullet.   We used an ultramicrotome to cut $\approx$100\,nm thick sections, which we placed onto Cu TEM grids with a 10\,nm amorphous carbon substrate (Ladd Research).   

We studied Andromeda using an FEI Titan transmission electron microscope (TEM) operated at 80 keV and 200 keV at the Molecular Foundry, Lawrence Berkeley National Laboratory (LBL).  
A 4-element Bruker silicon drift detector with a solid angle of 0.6 sr provided energy-dispersive X-ray spectroscopic (EDS) mapping in the Titan at count rates between 5 and 100 kcps.  
We used the Cliff-Lorimer approximation \citep{Cliff:1975vn} with a thin film correction as described in \cite{Gainsforth:2016im}.  
Most spectra had on the order of $10^6$ counts after background subtraction yielding detection limits $<$0.03 at\% for most elements (Tables \ref{tab:SulfideCompositions} -- \ref{tab:PyroxeneCompositions}).  
In the case of the large sulfide given by spectrum Sulfide\refspec{AndromedaSulfideBulk} in Table \ref{tab:SulfideCompositions} (abbreviated as: Table \ref{tab:SulfideCompositions}, Sulfide\refspec{AndromedaSulfideBulk}), a longer spectrum of 3$\times10^8$ counts allowed for quantification at hundreds of ppm concentration for several elements for which backgrounds were low.
For sensitive phases that lose Na during TEM analysis, we acquired a sequence of EDS maps and then extrapolated the original Na content based on the observed loss rate.
For the sulfide, the dose was sufficiently high that we applied a S volatilization correction  \citep{Gainsforth:2014iy}.
In many cases, it was not possible to fully remove the epoxy background from spectra and so we expect some O concentrations are a few percent too high.
For quantification of Fe and O in iron oxides, the excess O from epoxy could have significantly altered the conclusions, so we subtracted a nearby epoxy spectrum by normalizing to Cl when possible, or C when the Cl signal was too weak.

We compared compositions of the fine grained material (FGM) to Solar System abundances from \cite{Lodders:2003p2321}.
In most cases we chose to normalize against Mg rather than Si in order to the avoid possible confusion with SiO$_{2}$ dilution from aerogel.   To express elemental ratios with respect to protosolar values, we adopt the notation from astronomy:

\begin{equation}
    [A/B]_\odot = {\rm log}_{10} \left( \frac{X_A/X_B}{(X_A/X_B)_{\odot}} \right)
    \label{eq:protosolar}
\end{equation}

\noindent where $\odot$ indicates the accepted protosolar values from \cite{Lodders:2003p2321}.
Therefore, an $A/B$ ratio which is exactly protosolar would mean $[A/B]_\odot = 0$, a 10$\times$ enhancement in A relative to B gives $[A/B]_\odot = 1$ and a 3$\times$ depletion in A relative to B gives $[A/B]_\odot = -0.5$.

We carried out white-beam X-ray microdiffraction ($\mu$XRD) analyses on beamline 12.3.2 of the Advanced Light Source (ALS) at LBL \citep{Tamura:2009p21873}.   
The $\mu$XRD beam was generated by a superconducting bend magnet and then collimated to a spot size of approximately 0.5\,$\mu$m $\times$ 0.5\,$\mu$m with Kirkpatrick-Baez mirrors and slits.  
We recorded the XRD patterns on a Dectris Pilatus 1M camera.
By acquiring 2D spatial maps with a diffraction pattern at each position, we were able to analyze the crystal distortion as a function of its position.
By tracking the presence and location of a single diffracted beam in each of the diffraction patterns, we were able to produce topographs: images of the crystal locations and distortions much like dark-field electron microscopy does in a TEM.  For details on the method in general, see \cite{Black:2004p16776}, and for details on our specific implementation see \cite{Tamura:2014tq} and \cite{Gainsforth:2013gna}.

For estimation of gas temperatures during capture in aerogel, 
we computed dissociation energies for aerogel and troilite into gases using Density Functional Theory (DFT) implemented in the plane wave formalism as part of the Quantum Espresso suite \citep{Giannozzi:2009hx}.
We used Perdew-Burke-Ernzerhof pseudopotentials from the Standard Solid State Pseudopotentials Efficiency database \citep{1996PhRvL..77.3865P,Corso:2014gi,Vanderbilt:1990is,Prandini:PhlUrz_l}.
We used single-unit-cell quartz and troilite structures for solid SiO$_{2}$ and FeS, and a 5 {\AA} cubic volume for molecules.
Quartz was chosen as a proxy for aerogel since the two are expected to have formation energies within a few hundred meV of each other based on examination of SiO$_{2}$ structures in the Materials Project \citep{Jain:2013ku}, and examination of thermodynamic constants from the NIST-JANAF thermodynamic tables \citep{Chase:1998vz}.
Convergence criteria were set to $\Delta$E $\leq$ 40 meV in accordance with the expected accuracy of the pseudopotentials \citep{Lejaeghere:2016jx}.
Computation was done on the Vulcan, Nano and Etna compute clusters at LBL.

\section*{RESULTS}

\subsection*{Large Sulfide}

$\mu$XRD analysis showed that the sulfide was a polygonalized crystal,  consisting of three domains rotated by 0.35$^{\circ}$ from each other.
Fig. \ref{fig:TrackOptical}C shows a topograph produced from a Laue map that shows the location and size of each domain.
The reflections seen in XRD show multiple maxima within each reflection (expected for polygonalized crystals) and also show broadening of the peaks that indicate large internal stresses on the scale of 1$^{\circ}$ over the entire width of the crystal.
The diffraction pattern fits pyrrhotite 4C.
Because the internal strains are so large, it is not possible to differentiate troilite from pyrrhotite on the basis of the unit cell shape.

\begin{figure}[t!]
  \centering
  \includegraphics[width=\ImageWidthFactor\textwidth]{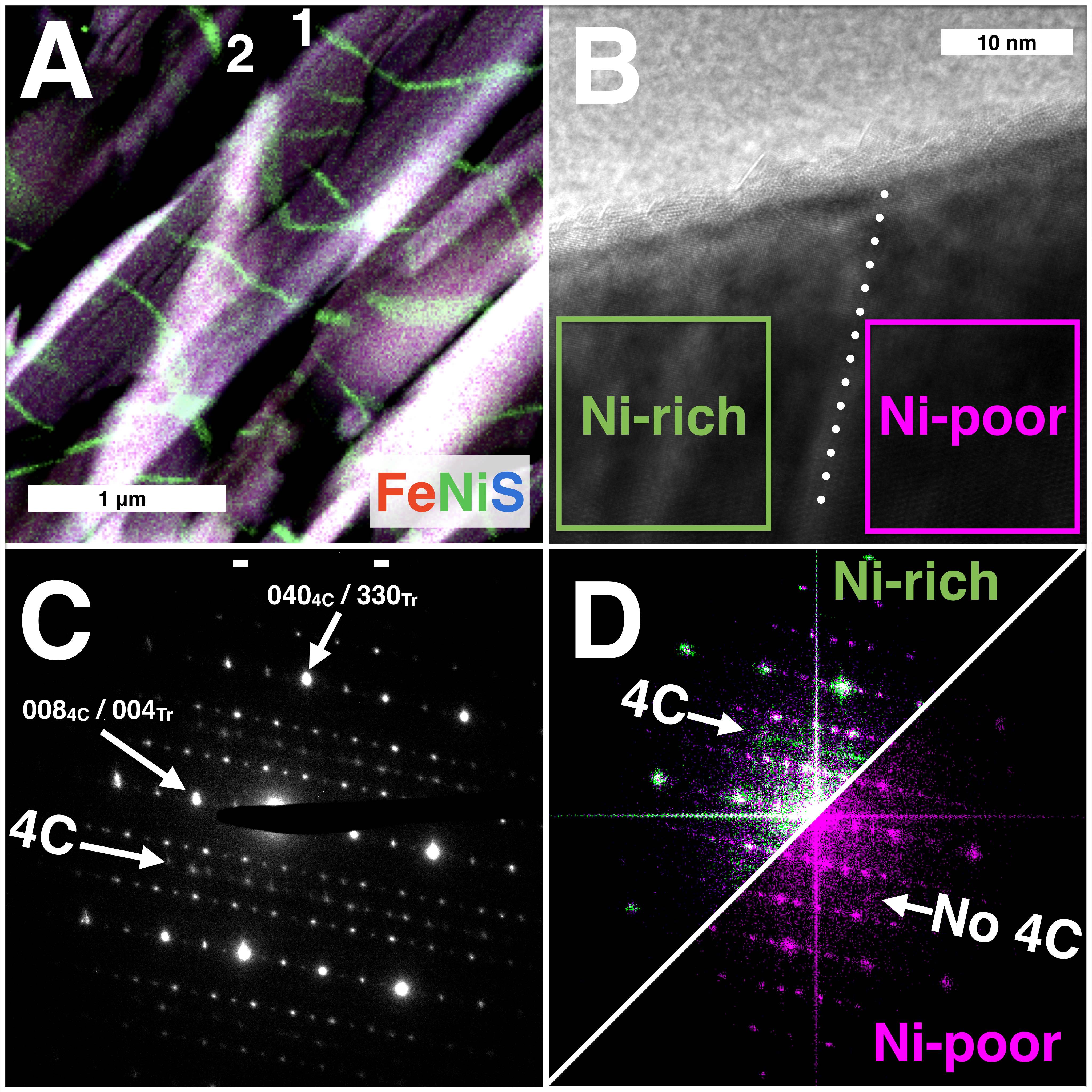}
  \caption{Figure \ref{fig:SulfideBands}.  A) EDS map showing troilite (purple) with Ni-rich bands (green).  B) HRTEM image down the 110 troilite axis with a Ni-rich band on the left and a Ni-poor region on the right.  The boxes show the regions used to produce FFTs of each region (shown in D).  C) SAED of region including Ni-rich and poor regions down the 110 (troilite) or 100 (pyrrhotite 4C) zone axis.  Pyrrhotite 4C superlattice reflections (4C) are marked alongside troilite (Tr). D) HRTEM FFT shows the same pattern.  Purple spots are from the FFT of the Ni-poor region (box shown in B) and green shows the FFT from the Ni-rich region (box shown in B).  The Ni-rich FFT shows the pyrrhotite 4C superlattice reflections. }
  \label{fig:SulfideBands}
\end{figure}

\begin{figure}[tbhp!]
  \centering
  \includegraphics[width=\ImageWidthFactor\textwidth]{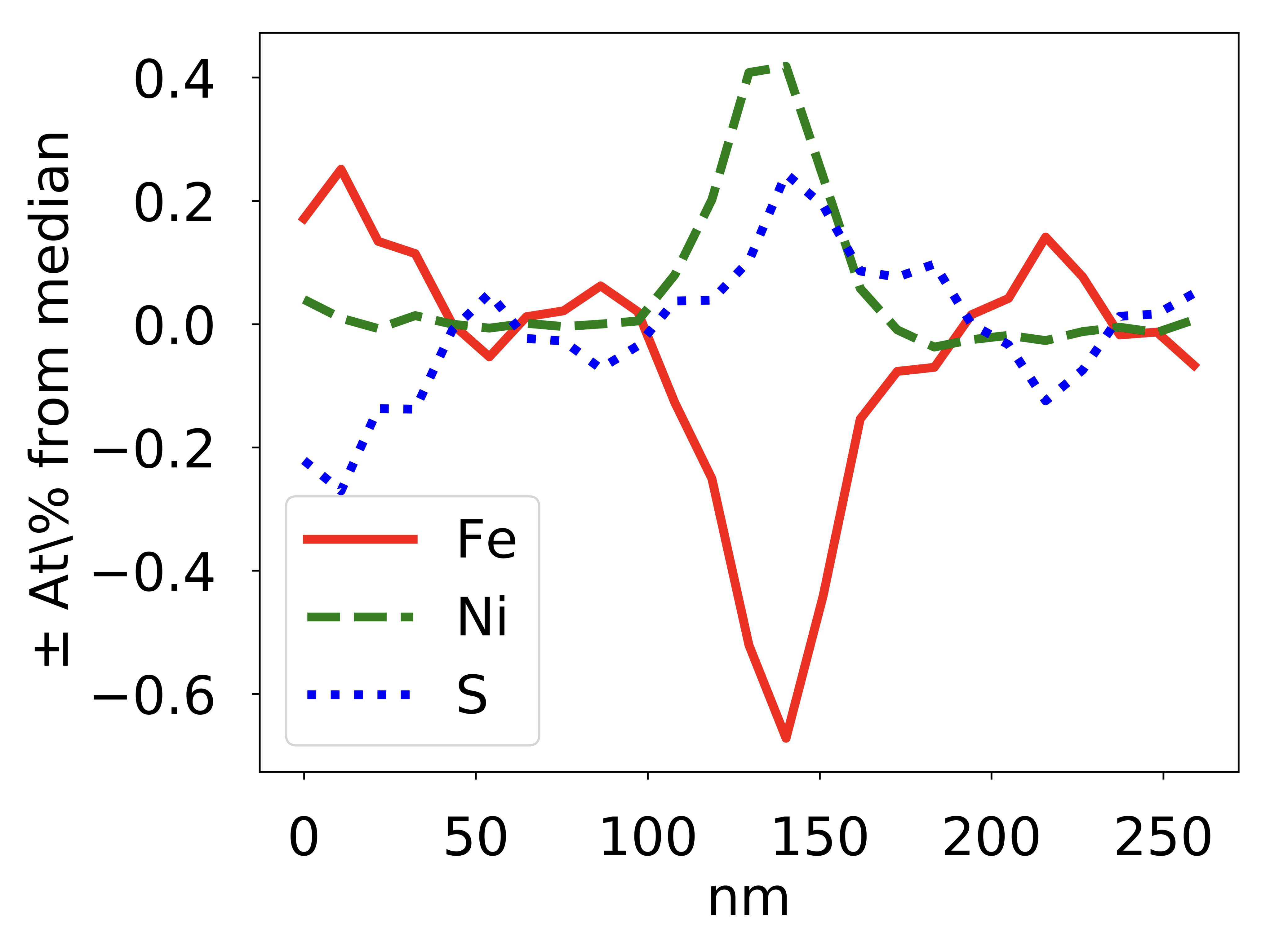}
  \caption{Figure \ref{fig:SulfideNiProfile}.  Average of three linescans across Ni-rich bands.  The Ni content is higher and the Fe content lower in the Ni-rich bands.  The variability of S across bands is not clear.}
  \label{fig:SulfideNiProfile}
\end{figure}

The (Fe+Ni)/S ratio measured by TEM/EDS was very close to 1:1, as expected for troilite, not pyrrhotite 4C (Table \ref{tab:SulfideCompositions}).   
Ni was strongly depleted with respect to Fe: [Ni/Fe]$_\odot<-1$.   Such Ni depletion is also seen in several other Stardust sulfides \citep{Joswiak2012:thetome, Gainsforth:2013vt}.
[S/Se]$_\odot$ = 0.02, indicating no significant fractionation of selenium from the protosolar abundance (S/Se = 6.8$\cdot 10^3$).   
High-resolution EDS maps showed that Ni is concentrated in bands no more than a few tens of nm thick rather than homogeneously distributed throughout the sulfide (Fig. \ref{fig:SulfideBands}A).   
The observed variability of Ni-rich band widths may be consistent with varying viewing angles of thin ($\leq 30$\,nm) lamellae  within the Ni-poor sulfide matrix.
The average of three linescans across Ni-rich bands is shown in Fig. \ref{fig:SulfideNiProfile} as the difference in composition from ideal troilite.
Ni and Fe are anti-correlated in the Ni-rich region.  
S appears to increase in the Ni-rich bands, but the deviation is no more than the deviation in S content outside the Ni-rich region and could be due to systematic error due to variable thickness of the sulfide.
The linescan is limited by the spatial resolution of the EDS map at $\approx$30 nm.

Selected Area Electron Diffraction (SAED) showed that the structure is best described by a superposition of troilite (zone 110) and pyrrhotite 4C (zone 100), though the pyrrhotite superlattice reflections are weak in comparison to those of the troilite (Fig. \ref{fig:SulfideBands}C).
HRTEM of the Ni-rich bands down the 110 zone axis of troilite show that they exhibit the pyrrhotite 4C superlattice reflection, and the Ni-poor regions do not (Fig. \ref{fig:SulfideBands}D).
We measured the broadening of the 11n family of superlattice reflections after subtracting  the instrumental response (22n) using the method of \cite{Gainsforth:2017er} and we found the spread to be 0.28 nm$^{-1}$ in the c* direction. 

\begin{table*}[h]
\scriptsize
\caption{\label{tab:SulfideCompositions}}
\setlength\extrarowheight{1.7pt}
\centering{}
\begin{tabular}{llllllllllllllll}
\multicolumn{16}{l}{{Table \ref{tab:SulfideCompositions}. Sulfide compositions from TEM EDS (atomic \%).}}\\

\hline 

 & \multicolumn{3}{c}{Atomic \%} &  & \multicolumn{3}{c}{ppm} &  &  &  &  &\\
 
\cline{2-4} \cline{6-8}

Spectrum                                 & S      & Fe     & Ni    & & Cr  & Mn   & Se  & $\Sigma$Cations & $\tau$ & Diffraction  & [Ni/Fe]$_\odot^a$ & [S/Se]$_\odot^a$ & Notes\\
 & & & & & & & & $\overline{\mathrm{(S+Se)}}$ & (nm g/cm$^3$) & &  & &\\
\\

\hline 

\multicolumn{16}{l}{Andromeda:}\\
Sulfide\refspec{AndromedaSulfideThinBlade}      & 49.86 & 49.99 & 0.15 & &     &     &     & 1.01            & 0     &               & -1.3   &       & Cliff-Lorimer\\
Sulfide\refspec{AndromedaSulfideBulk}           & 49.99 & 49.87 & 0.10 & & 350 &     & 70  & 1.00            & 279   &               & -1.5   & 0.02  & High-counts\\
Sulfide\refspec{AndromedaSulfideFeMatrix}       & 49.95 & 49.93 & 0.13 & &     &     &     & 1.00            & 358   & Troilite$^{b}$  & -1.3   &       & Fe matrix\\
Sulfide\refspec{AndromedaSulfideNiBand}         & 50.00 & 49.52 & 0.48 & &     &     &     & 1.00            & 358   & Pyrrhotite$^{b}$& -0.8   &       & Ni band\\
Sulfide\refspec{AndromedaSulfideNiBand2}        & 49.78 & 49.25 & 0.97 & &     &     &     & 1.01            & 358   &                 & -0.5   &       & Highest Ni\\
Sulfide\refspec{AndromedaIdealPyrrhotite}       & 53.35 & 39.73 & 6.93 & &     &     &     & 0.88            & n/a   &                 &        &       & Calculated$^{c}$\\
Sulfide\refspec{CPXSulfide}                     & 47.14 & 52.12 & 0.60 & &     &     &     & 1.12            & 300   &                 & -0.7   &       & P detected\\

\hline
\multicolumn{16}{l}{Febo:}\\
Sulfide\refspec{FeboSulfide}        			    & 53.34 & 45.76 & 0.79 & & 280 & 620 & 160 & 0.87            & 450   & nC            & -0.5   & -0.3  & \\

\hline 

\multicolumn{16}{l}{$^{a}$Ratio normalized to Solar System abundances, see text.}\\
\multicolumn{16}{l}{$^{b}$The thickness correction has been chosen to optimize for troilite.}\\[2pt]
\multicolumn{16}{l}{$^{c}$This idealized composition of the pyrrhotite bands was calculated in the discussion by combining HRTEM, diffraction and EDS.}\\[2pt]

\end{tabular}
\end{table*}

\subsection*{Fine-grained Material}

{
\begin{table*}[h]
\tiny
\caption{\label{tab:NanophaseCompositions}}
\setlength\extrarowheight{1.7pt}
\centering{}
\begin{tabular}{llllllllllllllllllll}
\multicolumn{19}{l}{{Table \ref{tab:NanophaseCompositions}. Nanophase object compositions from TEM EDS (atomic \%).}}\\

\hline 

& \multicolumn{14}{c}{Atomic \%} & & & &\\
 
\cline{2-15}

\# &      O &    Na &     Mg &    Al &     Si &      P &      S &     K &    Ca &    Ti &    Cr &     Mn &     Fe &    Ni &  $\tau$ &  O/Si &  \thead[l]{(Mg+Al+\\ Ca+Fe)\\ /Si} &                   Notes \\
\hline
 FGM\refspec{FGMBulk} &  52.92 &  0.27 &  1.33 &  0.25 &  17.39 &  0.17 &  11.60 &  0.13 &  0.57 &  b.d. &  0.04 &  0.03 &  15.17 &  0.14 &  300 &  3.04 &  1.00 &  FGM Bulk \\
 FGM\refspec{GEMS1} &  61.29 &  0.12 &  10.86 &  1.20 &  17.03 &  0.10 &  3.47 &  0.16 &  0.49 &  0.03 &  0.14 &  0.06 &  4.82 &  0.23 &  300 &  3.60 &  1.02 &      \\
 FGM\refspec{EA1} &  64.49 &  0.41 &  13.77 &  1.15 &  14.76 &  0.15 &  0.37 &  0.19 &  0.49 &  0.03 &  0.16 &  0.15 &  3.73 &  0.15 &  300 &  4.37 &  1.30 &      \\
 FGM\refspec{EA2} &  59.59 &  0.32 &  16.91 &  1.20 &  16.31 &  0.20 &  0.46 &  0.17 &  0.22 &  0.03 &  1.13 &  0.15 &  3.29 &  0.03 &  300 &  3.65 &  1.33 &      \\
 FGM\refspec{Alt1} &  43.52 &  0.04 &  4.24 &  0.72 &  13.98 &  b.d. &  17.51 &  0.09 &  0.37 &  0.09 &  0.06 &  19.11 &  0.25 &  0.03 &  300 &  3.11 &  0.40 &  Zn det. \\
 FGM\refspec{Alt2} &  66.17 &  0.80 &  14.70 &  1.19 &  12.73 &  0.22 &  0.93 &  0.07 &  0.26 &  0.05 &  0.09 &  b.d. &  2.55 &  0.19 &  300 &  5.20 &  1.47 &      \\
 FGM\refspec{Alt3} &  62.90 &  b.d. &  12.41 &  0.93 &  12.95 &  0.12 &  0.80 &  0.07 &  0.58 &  0.04 &  0.21 &  0.09 &  8.62 &  0.29 &  300 &  4.86 &  1.74 &      \\
 FGM\refspec{Alt4} &  60.55 &  0.38 &  9.04 &  1.27 &  16.14 &  0.12 &  2.68 &  0.10 &  1.31 &  0.02 &  0.16 &  0.11 &  7.93 &  0.22 &  300 &  3.75 &  1.21 &  Has Fe-Oxides \\
 FGM\refspec{Alt5} &  62.16 &  0.19 &  10.74 &  0.77 &  14.48 &  0.44 &  0.73 &  0.12 &  0.86 &  0.01 &  0.42 &  0.10 &  8.95 &  0.02 &  300 &  4.29 &  1.47 &  Has Fe-Oxides \\
 FGM\refspec{Slag1} &  62.37 &  0.27 &  6.29 &  1.31 &  18.10 &  0.08 &  1.46 &  0.14 &  1.72 &  0.03 &  0.12 &  0.04 &  7.74 &  0.34 &  300 &  3.45 &  0.94 &      \\
 FGM\refspec{Slag2} &  57.64 &  0.14 &  3.56 &  0.27 &  20.31 &  0.039 &  7.65 &  0.03 &  0.21 &  b.d. &  0.04 &  0.06 &  9.90 &  0.15 &  300 &  2.84 &  0.69 &      \\
 FGM\refspec{Slag3} &  65.24 &  0.41 &  9.51 &  0.89 &  13.42 &  0.25 &  1.07 &  0.21 &  0.94 &  0.02 &  0.15 &  b.d. &  7.81 &  0.06 &  300 &  4.86 &  1.43 &  Zn det., Has Fe-Oxides \\
 FGM\refspec{Slag4} &  58.75 &  0.07 &  2.60 &  1.62 &  21.53 &  0.30 &  1.33 &  0.38 &  0.38 &  0.06 &  0.32 &  b.d. &  12.08 &  0.61 &  300 &  2.73 &  0.77 &      \\


\end{tabular}
\end{table*}

}

Andromeda contained a field of FGM, partially embayed in the large sulfide,  with $\approx$50 Mg-rich particles in the ultramicrotomed section shown in Fig. \ref{fig:AndromedaPanorama}.  
We estimate that several hundred such particles were probably present in total, considering that the microtomed section is only a 100-200 nm thick sampling of a 3D volume.  
TEM and STXM analysis showed that the embedding epoxy penetrated the interstial spaces in the FGM during sample preparation, indicating that the material was porous prior to embedding. 

When normalized to Si and protosolar ratios, we find that the FGM is depleted in Na, Mg, Al, Ca, Cr, Mn and Ni, and enriched in P, S, Fe and K (Table \ref{tab:NanophaseCompositions}, FGM\refspec{FGMBulk}).
If the FGM had protosolar elemental abundances before capture, we calculated an upper limit of 1\% contamination by SiO$_2$ and 30\% by FeS.
K was present at $\approx$2 times protosolar abundance, normalized to Si, but no K-rich minerals were observed.  

We studied five classes of particles in the FGM field, discussed in detail below:  GEMS-like objects, EA-like objects, Fe oxides, kosmochloric pyroxene and sulfides.

\subsubsection*{GEMS-like objects}

The mean composition of GEMS differs from that of the Solar System \citep{Keller:2011bx}.  
\cite{Keller:2011bx} and \cite{Messenger:2015gz} measured compositions of 287 GEMS, from which we calculated mean compositions and standard deviations for each element.  We used a $\chi^2$ test to determine if objects within the FGM were consistent in their elemental composition with GEMS:

\begin{equation}
 \chi^2 = \sum_{\rm elements} { (X - \bar{X}_{\rm GEMS})^2 \over {\sigma_{\rm GEMS}^2} } 
\end{equation}

where $X$ is the atomic fraction of an element in the particle, $\bar{X}_{\rm GEMS} $ and $\sigma_{\rm GEMS}$ are the mean and standard deviation respectively of the atomic elemental fractions, measured in a large number of GEMS from \cite{Keller:2011bx} and \cite{Messenger:2015gz}.
To compute $\chi^2$ we used atomic fractions of eight elements and report the reduced $\chi^2$ (${\chi}_\nu^2$, $\nu = 7$ degrees of freedom) along with $p$, the probability that the composition is inconsistent with that of GEMS distribution.
Since low-Z elements are suceptible to significant instrumental differences and are often more volatile, we have not included elements with Z $\le$ Na in the computation of $\chi^2$.
For the purpose of computing $p$, we made the simplifying assumption that the elemental abundances are normally distributed independent variables (e.g., we ignored the correlation between Fe and S).   
This assumption is conservative in the sense that it tends to overestimate $p$.

GEMS and some EAs have compositions similar to each other and distinct from other materials.
As can be seen in Table \ref{tab:NanophaseCompositions} we acquired quality EDS maps and images of about 10 objects within the FGM that have compositions close to GEMS.
To estimate bias in our selection we used k-means clustering \citep{Arthur:2007tv} on the EDS map from Fig. \ref{fig:AndromedaPanorama} to identify 52 regions which could be GEMS-like, so we studied $\sim$20\% of the GEMS-like objects in this field. 

\begin{figure}[t!]
  \centering
  \includegraphics[width=\ImageWidthFactor\textwidth]{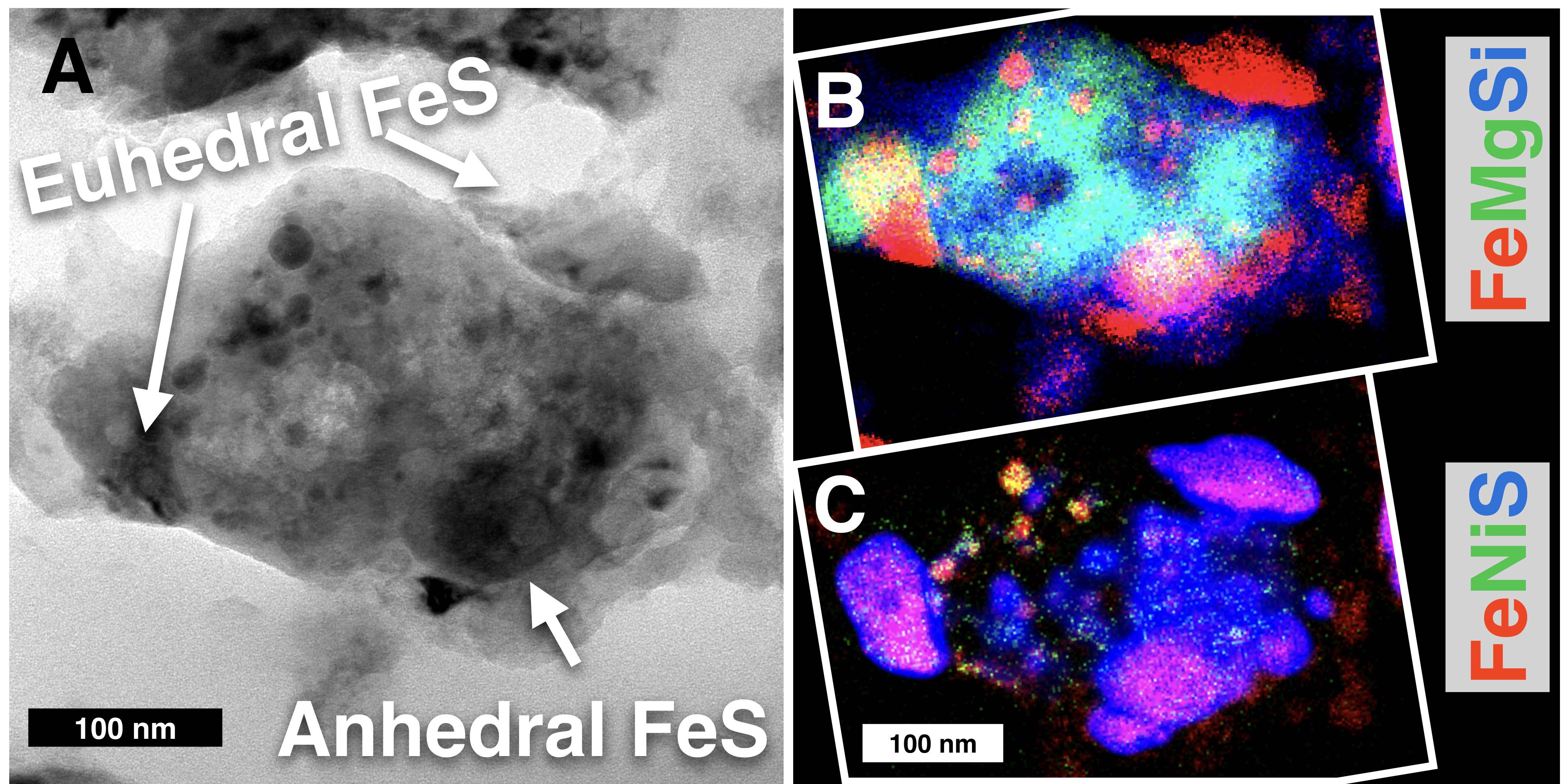}
  \caption{Figure \ref{fig:GEMS}.  A) BF TEM image of FGM\refspec{GEMS1}, a GEMS-like object in Andromeda's FGM.  Euhedral sulfides are present on one side of the object.  B) EDS map showing Fe(red), Mg (green), Si (blue).  Mg shows distinct inhomogeneity.  C) EDS map of showing Fe (red), Ni (green), S (blue, with a three pixel gaussian filter).  Fe-Ni metals are present within the body of the object.}
  \label{fig:GEMS}
\end{figure}

We observed an object (FGM\refspec{GEMS1}) which had the appearance and composition of a GEMS (Fig. \ref{fig:GEMS}, Table \ref{tab:NanophaseCompositions}, FGM\refspec{GEMS1}, $p$ = 0.2).
The top and left edges were bounded by $\approx$100\,nm euhedral sulfides.   The core contained Fe-Ni metal inclusions 5-20 nm in size.
Within the metal inclusions, [Ni/Fe]$_\odot$ was $\approx$ 0.3,  while the external sulfide had [Ni/Fe]$_\odot \approx$ 0.0.
EDS maps show zoning of Mg and Ca within the core, a feature that has been noted within GEMS in IDPs \citep{Keller:2011bx, Joswiak:1996vv}.  
The sulfide on the lower right was more rounded and some sulfur was present in the nearby silicate portion, penetrating approximately half of the silicate.
One metal-core/sulfide-shell structure was present in the silicate region containing the diffuse S.

\subsubsection*{EA-like objects}

\begin{figure}[t]
  \centering
  \includegraphics[width=\ImageWidthFactor\textwidth]{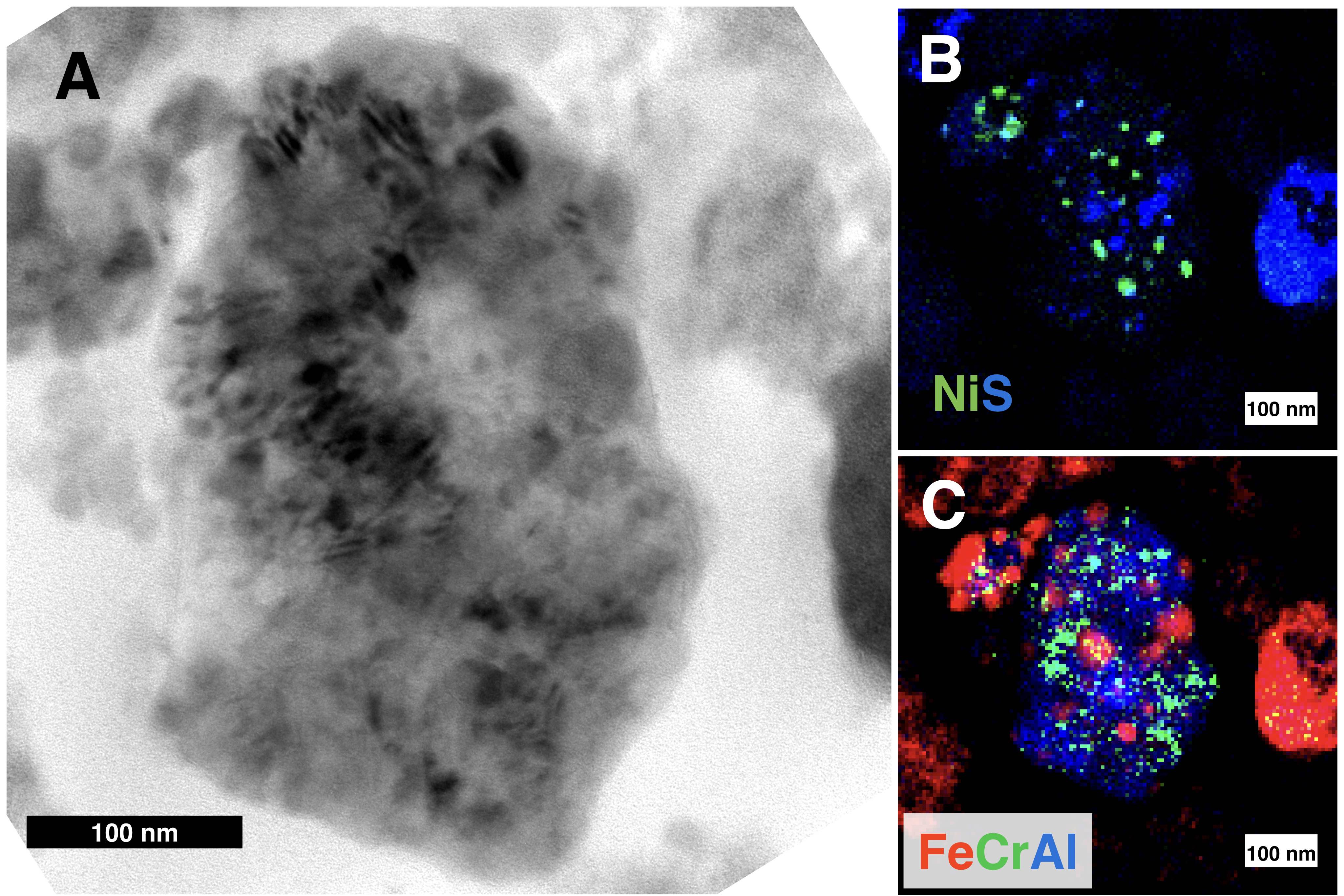}
  \caption{Figure \ref{fig:EquilibratedAggregate}.  A) BF TEM image of FGM\refspec{EA1}, an equilibrated aggregate (EA) in Andromeda's FGM.  B) EDS map of EA showing the location of Ni-rich metals (green) and the presence of sulfides (blue).  C) EDS Map of EA showing the location of Cr-rich inclusions (green) and an Al-rich inclusion and surrounding Al-rich silicates (blue).}
  \label{fig:EquilibratedAggregate}
\end{figure}

Two objects in the FGM field showed morphologies reminiscent of equilibrated aggregates in CP-IDPs.  
FGM\refspec{EA1} (Fig. \ref{fig:EquilibratedAggregate}) contained Mg-rich $\approx$ 10\,nm silicate crystals. 
The silicate crystals showed mosaicity, that is, they were oriented within a few degrees of each other.  
It also contained Fe-Ni metals, sulfides, Cr-rich hotspots (probably chromite), and Al-rich inclusions.
The outline of the object was defined by the euhedral edges of the silicate crystals.  
There was no attached silica melt, anhedral sulfide, or any other phase to indicate alteration during capture.

FGM\refspec{EA2} was a second particle with an equilibrated aggregate morphology and contained Mg-rich silicate crystals.
There were several Fe-Ni metal and sulfide inclusions, and no core-rim structures.
The periphery was surrounded by euhedral chromite crystals, and was well defined by their shapes.

\subsubsection*{Heated GEMS-like and EA-like objects}

Additional objects were present that had characteristics of GEMS or EAs, but showed indications of significant heating.

FGM\refspec{Alt1} is morphologically reminiscent of GEMS but is compositionally inconsistent with GEMS ($p$ > 0.99) because of a large excess of sulfide and Cr relative to the other elements.
In the core are Fe-Ni metal inclusions with [Ni/Fe]$_\odot$ = -0.04.
There was one euhedral sulfide on the left face with [Ni/Fe]$_\odot$  = -0.55.
The remaining sulfide resided around the periphery, was morphologically nebulous, and conformally coated the object.
A metal-core/sulfide-rim structure was visible in the silicate portion of the object.

\begin{figure}[t!]
  \centering
  \includegraphics[width=\ImageWidthFactor\textwidth]{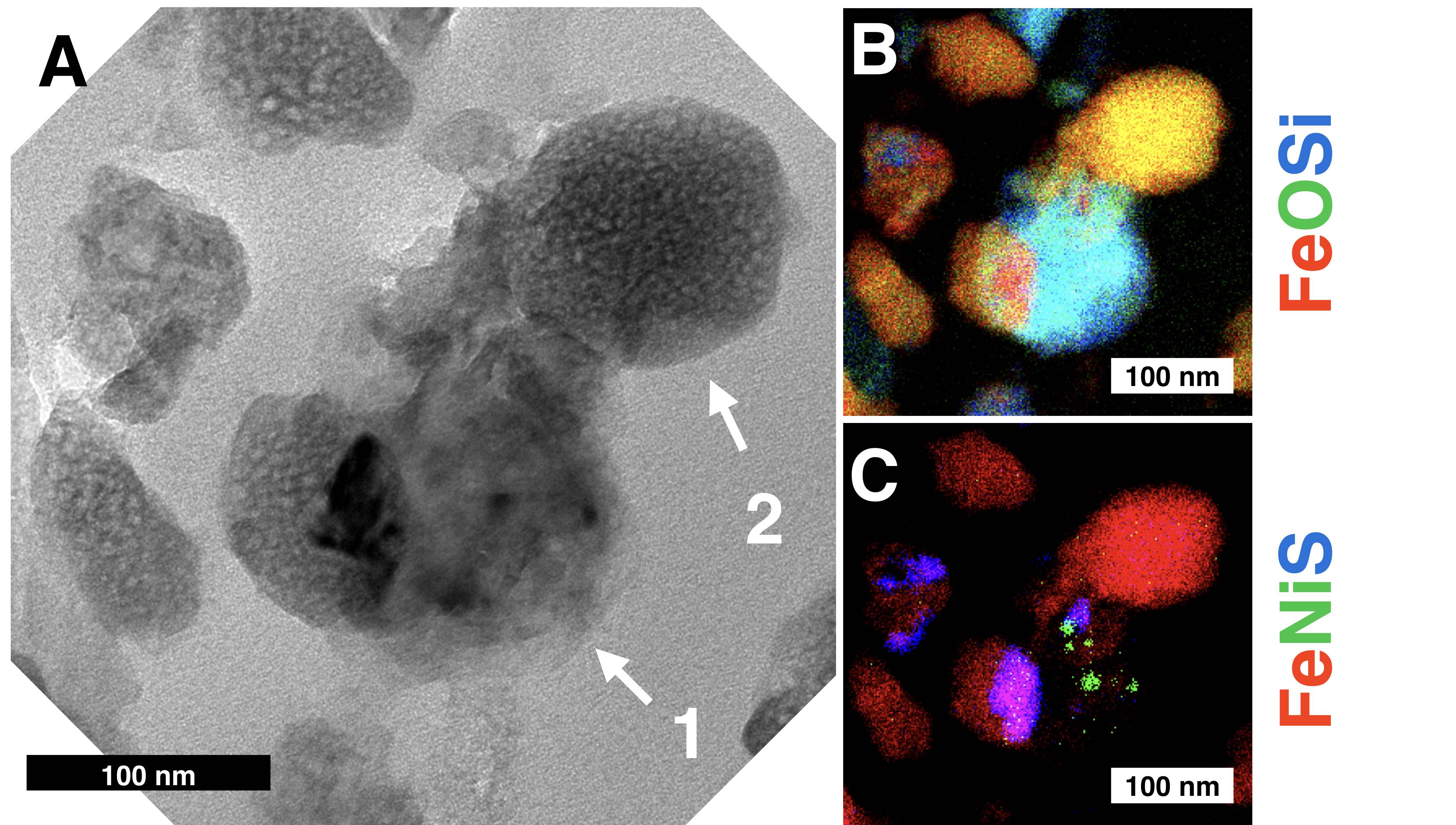}
  \caption{Figure \ref{fig:Alt4}.  A) BF image of FGM\refspec{Alt4}.  The location marked (1) is the equilibrated aggregate.  (2) is one of several Fe-oxides.   B) EDS map showing the Si-rich aggregate (cyan) contrasted with the Fe-oxides (yellow).  C) EDS map showing the presence of Ni-rich metals inside the equilibrated aggregate (green), the Fe-oxides (red) and sulfides (purple).}
  \label{fig:Alt4}
\end{figure}

FGM\refspec{Alt2}, FGM\refspec{Alt3} and FGM\refspec{Alt4} were all spherical objects  100 - 200\,nm across with $p$ = 0.5--0.8.  They contained Fe-Ni metals and sulfides.
They did not contain euhedral sulfides around the periphery, with the possible exception of one sulfide adjacent to FGM\refspec{Alt2}.
They did not show evidence for the core-rim structures caused by aerogel capture, and they did not show evidence for excess SiO$_2$.
FGM\refspec{Alt4} (Fig. \ref{fig:Alt4}) was apparently sintered to amorphous Fe-oxide. Additional Fe oxides containing Mg, Al, Si, P, S, Ca and Ni were present in the vicinity. These oxides are described in the next section and are a primary reason for considering these to be heated.

FGM\refspec{Alt5} contained crystalline Mg-rich silicates and Cr-rich inclusions.  It is  associated with an Fe-oxide similar to the oxide adjacent to FGM\refspec{Alt4}.  
Fe was also visible as a web-like structure weaving around the periphery of the object and outlining the crystals.  
There are no core-rim structures or excess silica.

FGM\refspec{Slag1} was reminiscent of GEMS but contained a metal-core/sulfide-rim structure.
There were no euhedral crystals around the periphery:  instead, all crystals within and without were rounded.
There was no evidence of crystalline silicate, but the Mg formed a web-like structure throughout the object interconnecting the Fe-Ni metals.
The outer periphery of the object was rich in Ca.

\begin{figure}[t]
  \centering
  \includegraphics[width=\ImageWidthFactor\textwidth]{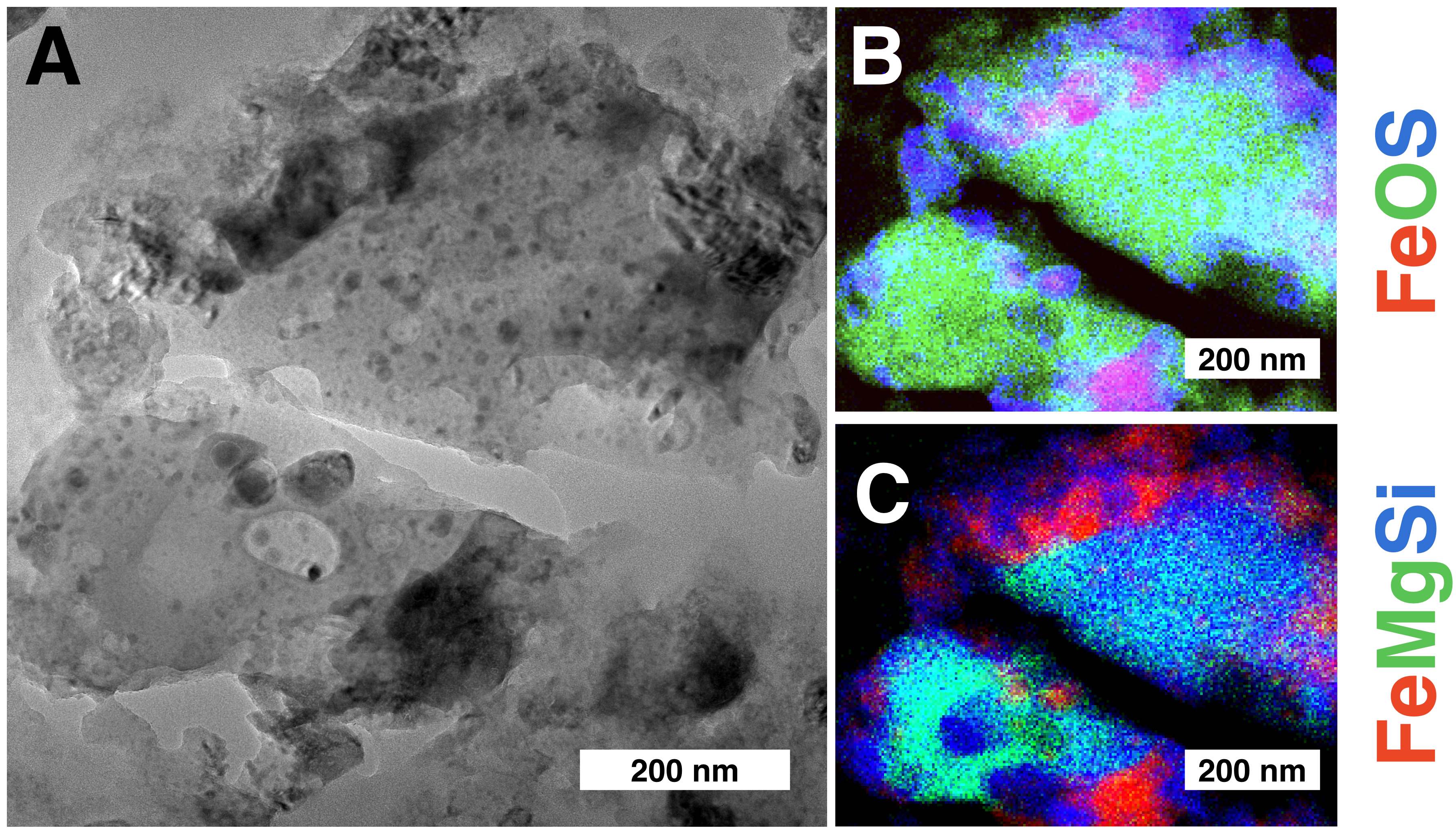}
  \caption{Figure \ref{fig:Slag2}.  A) BF image of FGM\refspec{Slag2}.  Vesiculated structure and Fe-core/sulfide-rim structures are visible. The periphery was surrounded by sulfides.  The object is bisected by ultramicrotomy.  B) EDS map showing Fe-cores (red) and sulfide rims (blue).  The periphery also contains sulfides which have a web-like texture from volatilization during aerogel capture.  C) EDS map showing vesicles in the silicate where Mg (green) is absent.}
  \label{fig:Slag2}
\end{figure}

FGM\refspec{Slag2} was similar to common aerogel capture products seen by \cite{Ishii:2008p2159} (Fig. \ref{fig:Slag2}).    
It contained multiple metal-core/sulfide-rim structures.  We observed  vesicular structures and substantial excess silica. The periphery was lined by amorphous web-like FeS.

FGM\refspec{Slag3} was located at the boundary between the sulfide and FGM, touching the sulfide, and contained amorphous and poorly-crystalline objects.  
The outlines of Al-rich silicate, Ca-rich silicate, and Mg-rich silicate could be seen but were not crystalline.  
An amorphous silicate with Fe-Ni metal grains and Fe-Ni-S grains was attached to one side.
Several small beam-sensitive, amorphous Fe-oxides were present around the periphery.

FGM\refspec{Slag4} was also located at the sulfide/FGM boundary and contained the highest K and P concentration of any particle we measured.  
The P was largely concentrated into a $\approx$ 10 nm diameter Fe-Ni-P hotspot that could have been schreibersite, and appeared to have a core-shell structure in HAADF.
Fe-Ni metals were also present, sulfur was diffused throughout the silicate, and no sulfides were visible.
The Mg and Al were present near the center but segregated from each other, and a 5 nm Ti hot spot was present in the Al-rich portion of the EDS map.

\subsubsection*{Iron oxides in FGM}

Diffraction analyses of the Fe oxides near FGM\refspec{Alt4} showed that they were amorphous.  
Comparison of HAADF images taken before and after EDS mapping of the oxides around FGM\refspec{Slag3} showed that the Fe oxides were beam-sensitive.

The Fe oxides were most prevalent at the interface between the FGM and the sulfide, and in the large voids between clumps of material within the FGM.
The oxides were also present on the leading edge of the sulfide away from the FGM.
They were not frequently found within clumps of FGM material.

\begin{figure}[t]
  \centering
  \includegraphics[width=\ImageWidthFactor\textwidth]{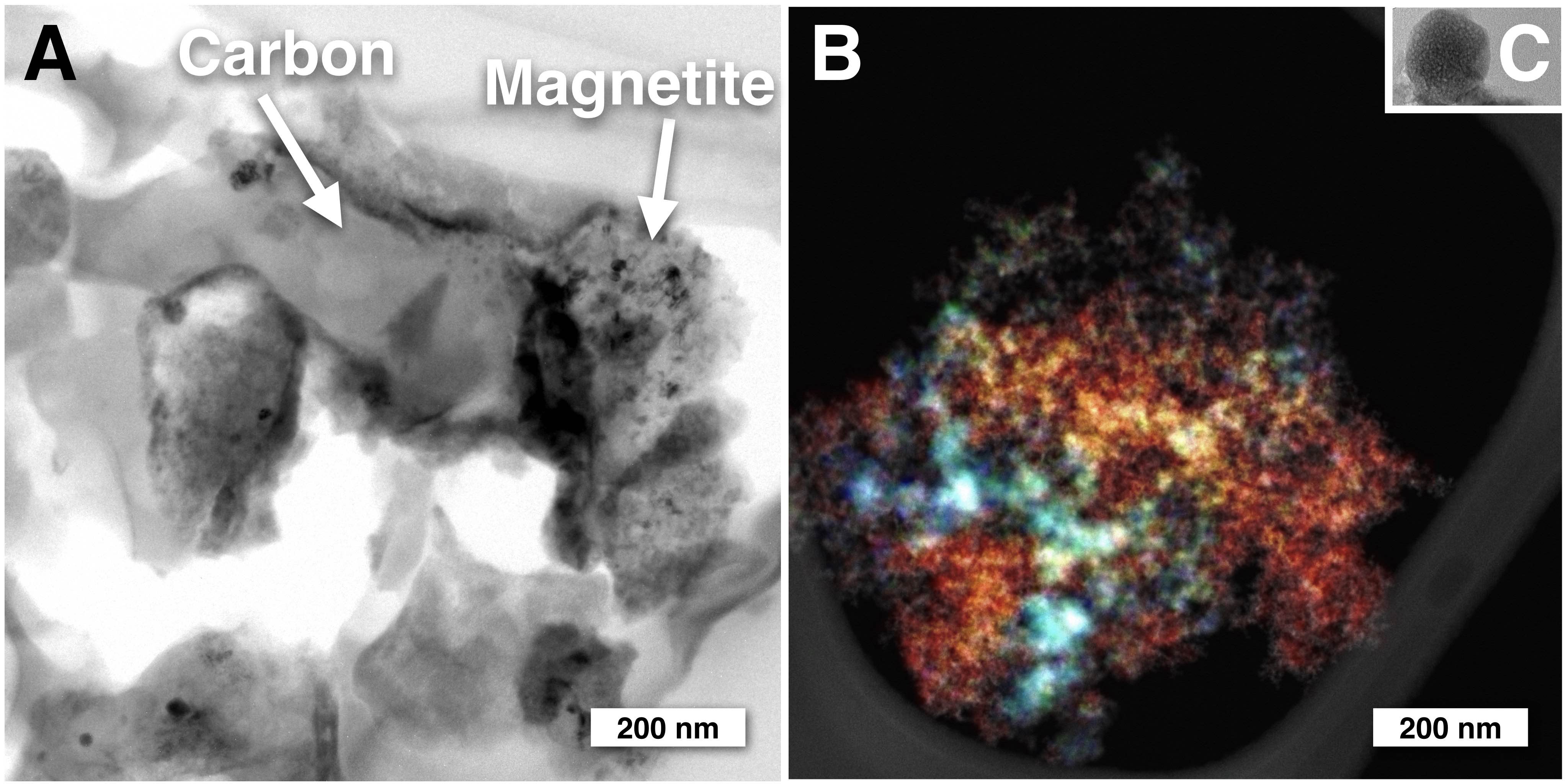}
  \caption{Figure \ref{fig:FeOxideAnalogs}.  A) BF image of magnetite from the IDP Lambda.  B) LRGB image of Fe-smoke made by \cite{JosephANuth:SmokeSynthesis}.  Luminance channel is HAADF, Fe (red), O (green), Si (blue).  C) An Fe-oxide from FGM\refspec{Alt4} to scale for comparison.  All scale bars are 200 nm.}
  \label{fig:FeOxideAnalogs}
\end{figure}

For comparison, we examined a magnetite rim from a lightly heated CP-IDP (Lambda, Cluster IDP L2071,17).  
The bright-field image shown in Fig. \ref{fig:FeOxideAnalogs}A shows that the rim was nanocrystalline, although it may also have included some amorphous regions.  
The EDS spectrum in Table \ref{tab:FeOxides} (FeOx\refspec{Lambda-Magnetite}) shows abundances of Mg, Al, P, S and Ni $\le$ 1 at\%.

We also studied Fe-O-Si smokes made by \cite{JosephANuth:SmokeSynthesis} shown in Fig. \ref{fig:FeOxideAnalogs}B.  
Here, amorphous and crystalline objects a few nm wide aggregated to produce smoke particles.  
The EDS overlay in the image shows that the Fe-O-Si elements are heterogeneously distributed, unlike in Andromeda oxides.

\begin{table*}[ht]
\scriptsize
\caption{\label{tab:FeOxides}}
\setlength\extrarowheight{1.7pt}
\centering{}
\begin{tabular}{llllllllllllllll}
\multicolumn{16}{l}{{Table \ref{tab:FeOxides}. Fe Oxide compositions from TEM EDS (atomic \%).}}\\
\hline 

& \multicolumn{11}{c}{Atomic \%}&\\
 
\cline{2-12}

\# &      O &    Na &    Mg &    Al &    Si &     P &     S &     K &    Ca &     Fe &    Ni &  $\tau$ &  O/Fe & Background$^1$ &                       Notes \\
\hline
 FeOx\refspec{Alt4-FeOxide1} &  65.65 &  b.d. &  b.d. &  0.15 &  0.68 &  det. &  1.64 &  b.d. &  0.34 &  31.49 &  0.06 &  300 &  2.08 &  Cl &  Near FGM\refspec{Alt4} \\
 FeOx\refspec{Alt4-FeOxide2} &  50.89 &  b.d. &  b.d. &  b.d. &  2.33 &  0.49 &  1.28 &  b.d. &  0.58 &  44.44 &  b.d. &  300 &  1.15 &  Cl &  Near FGM\refspec{Alt4} \\
 FeOx\refspec{Alt4-FeOxide3} &  63.88 &  b.d. &  b.d. &  b.d. &  4.29 &  0.69 &  2.23 &  b.d. &  0.63 &  28.28 &  b.d. &  300 &  2.26 &  C &  Near FGM\refspec{Alt4} \\
 FeOx\refspec{Alt4-FeOxide4} &  59.70 &  b.d. &  b.d. &  b.d. &  0.96 &  0.37 &  1.10 &  b.d. &  0.47 &  37.40 &  b.d. &  300 &  1.60 &  Cl &  Near FGM\refspec{Alt4} \\
 FeOx\refspec{Alt4-FeOxide5} &  56.70 &  b.d. &  0.61 &  b.d. &  2.42 &  0.37 &  1.47 &  b.d. &  0.43 &  38.00 &  b.d. &  300 &  1.49 &  Cl &  Near FGM\refspec{Alt4} \\
 FeOx\refspec{Alt5-FeOxide6} &  63.52 &  0.93 &  0.60 &  b.d. &  1.38 &  0.50 &  1.89 &  0.32 &  1.95 &  28.92 &  b.d. &  300 &  2.20 &  C &  Near FGM\refspec{Alt5} \\
 FeOx\refspec{Alt5-FeOxide7} &  59.40 &  1.46 &  b.d. &  b.d. &  1.67 &  0.96 &  2.60 &  0.68 &  2.16 &  31.07 &  b.d. &  300 &  1.91 &  C &  Near FGM\refspec{Alt5} \\
 FeOx\refspec{Lambda-Magnetite} &  53.11 &  b.d. &  0.41 &  0.18 &  b.d. &  0.20 &  0.69 &  b.d. &  b.d. &  44.53 &  0.83 &  300 &  1.19 &  Cl &  Lambda magnetite.  Cr det. \\

\multicolumn{11}{l}{$^{1}$Element for normalizing and subtracting the background.  See methods.}\\[2pt]

\end{tabular}
\end{table*}

\subsubsection*{Kosmochloric pyroxene}

We observed a kosmochloric pyroxene (i.e., Na-Cr coupled substitution) in the FGM similar to those commonly found in other Wild 2 samples and in CP-IDPs \citep{Joswiak:2009p21878}.
The elemental quantification is shown in Table \ref{tab:PyroxeneCompositions} Px\refspec{AndromedaCPX}.  
Small amounts of P and S in the spectrum may have been due to neighboring glassy material and neighboring sulfide.  Its diffraction was consistent with diopside, as should be expected based on its composition.  
The Na content was 0.159 cations per 6 oxygens, which was well balanced by Ti$^{4+}$, Al$^{3+}$ and Cr$^{3+}$: Al + Cr + 2Ti = 0.174 cations per 6 oxygens,  suggesting (Ti, Al, Cr)-Na coupled substitution.
A small sulfide-metal appendage was found on one side of the pyroxene (Table \ref{tab:SulfideCompositions}, Sulfide\refspec{CPXSulfide}). 
The Ni/Fe ratio was significantly larger than the primary sulfide impactor.

\begin{table*}[t]
\begin{centering}
\scriptsize
\begin{tabular}{lllllllllllll}
\multicolumn{13}{l}{Table \ref{tab:PyroxeneCompositions}. Pyroxene compositions from TEM EDS.}\\
\hline 
 & \multicolumn{8}{c}{Normalized oxide weight \%} &  & \\
\cline{2-11} 
Spectrum  & SiO$_{2}$ & TiO$_{2}$ & Al$_{2}$O$_{3}$ & Cr$_{2}$O$_{3}$ & MgO & CaO & FeO & Na$_{2}$O & P$_{2}$O$_{5}$  & S     & \multicolumn{2}{l}{Phase$^{b}$}\\
\hline 

Px\refspec{AndromedaCPX} & 56.14 & 0.25 & 0.95 & 4.23 & 14.81 & 16.26 & 3.94 & 2.28 & 1.00 & 0.14 & \multicolumn{2}{l}{En$_{40}$Fs$_{07}$Wo$_{33}$Ko$_{14}$Ja$_{06}$}\\
 &  &  &  &  &  &  &  &  &  & \\
 & \multicolumn{8}{c}{Cations per 6 oxygens} &  & \\

\cline{2-11} 

Spectrum               & Si    & Ti    & Al    & Cr    & Mg    & Ca    & Fe    & Na    & P     & S     & $\Sigma$Cations & $\tau^b$\\
\hline 

Px\refspec{AndromedaCPX} & 2.02 & 0.01 & 0.04 & 0.12 & 0.80 & 0.63 & 0.12 & 0.16 & 0.03 & 0.01 & 3.90           & 300 \\

\hline 
\multicolumn{13}{l}{$^{a}$En=enstatite, Fs=ferrosilite, Wo=wollastonite, Ko=kosmochlor, Ja=jadeite.}\\
\multicolumn{13}{l}{$^{b}$nm$\cdot$g/cm$^{3}$.}\\
\end{tabular}
\par\end{centering}{\scriptsize \par}

\protect\caption{\label{tab:PyroxeneCompositions}}
\end{table*}

\section*{DISCUSSION}

\subsection*{Wild 2 FGM compared to GEMS}

Hypervelocity capture of particles in aerogel can produce objects that are morphologically reminiscent of GEMS \citep{2008M&PS...43...97L, Ishii:2008p2159}.  
Four common signatures distinguish these objects from GEMS.  
1) Excess silica is usually present due to mixing with the aerogel.  In some cases the original compositions are still preserved except for the excess SiO$_{2}$.  
2) The rapid heating of sulfides reduces them to iron metal and subsequent rapid cooling recondenses the sulfur on the outside to form sulfide rims on the iron cores \citep{Leroux:2008p24341}.
3) Vesicular structures from rapid volatilization are often observed.
4) Euhedral shapes on crystals, especially sulfides, are lost as they are liquified or volatilized.

\cite{Keller:2011bx} studied GEMS grains and noted that they had varying degrees of silicate polymerization.
They hypothesized that this was related to nebular interactions with hydrogen and water which resulted in excess oxygen and OH within the silicate matrix.

\begin{figure}[hbt]
  \centering
  \includegraphics[width=\ImageWidthFactor\textwidth]{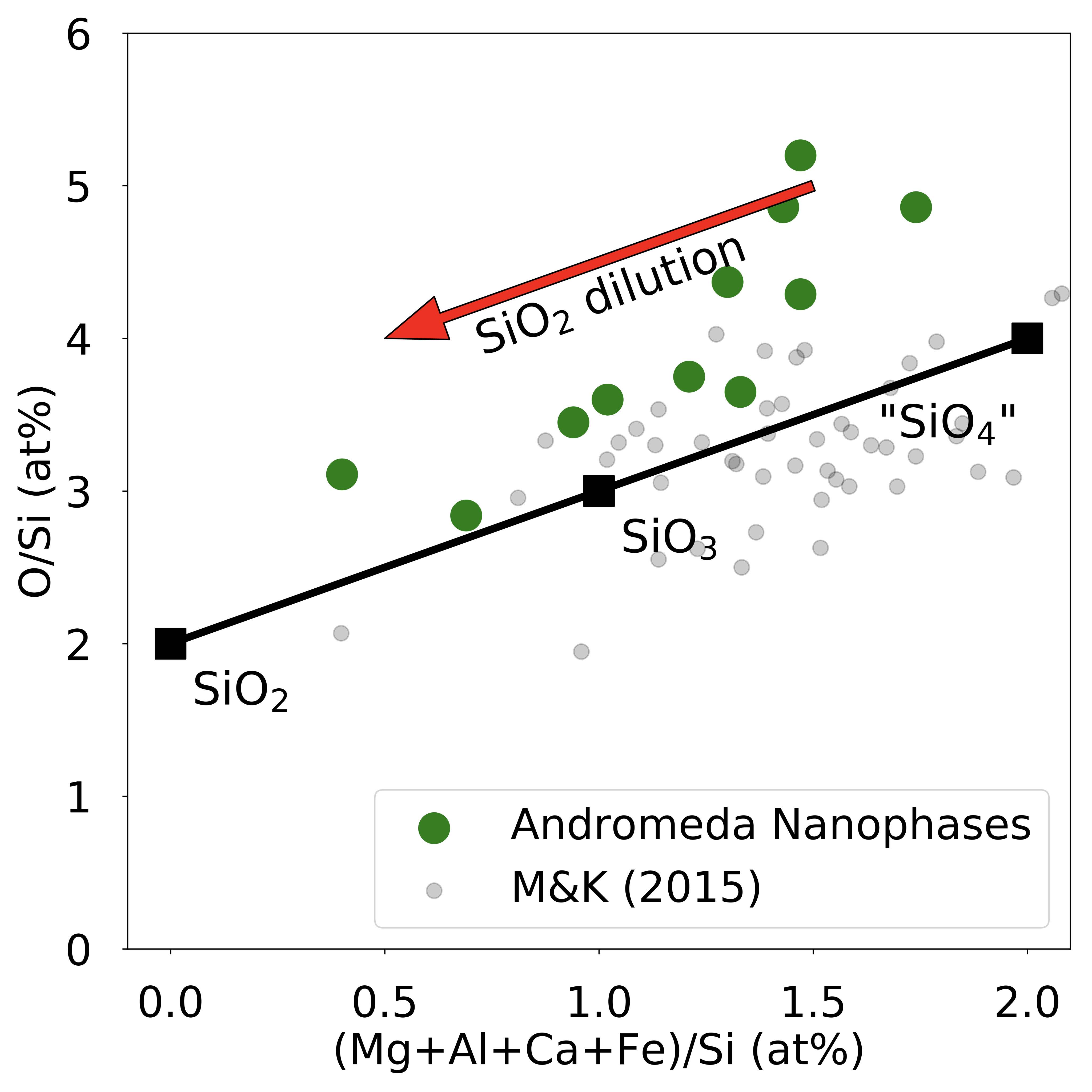}
  \caption{Figure \ref{fig:GEMSPolymerization}.  The silicate polymerization of nanophase objects from Andromeda's FGM is plotted after the style of \cite{Keller:2011bx}.  The nanophase objects include the GEMS and EA objects as well as the thermally modified objects, yet only three of the studied Andromeda nanophase objects plot outside the field between SiO$_3$ and SiO$_4$ as might be caused by aerogel dilution (shown by red arrow).  Points above the line show excess oxygen, and points below the line are deficient in oxygen relative to expected stoichiometries.  Gray points are from a recent collection of IDPs not contaminated with silicone oil \citep{Messenger:2015gz}.}
  \label{fig:GEMSPolymerization}
\end{figure}

We plot the compositions of the GEMS-like and EA-like objects  in Fig. \ref{fig:GEMSPolymerization} in a fashion modeled after \cite{Keller:2011bx}, their Fig. 7.
Most of our objects show excess O which is likely an artifact of the surrounding epoxy,  so this analysis is not sensitive to the presence of OH.
However, we still find a very similar trend to that observed by \cite{Keller:2011bx}, namely these objects have a variable silicate polymerization between SiO$_{3}$ and SiO$_{4}$.
Aerogel dilution would push the composition towards SiO$_{2}$, but the distribution of our objects does not favor compositions more silica-rich than SiO$_{3}$.

\begin{figure}[t!]
  \centering
  \includegraphics[width=\ImageWidthFactor\textwidth]{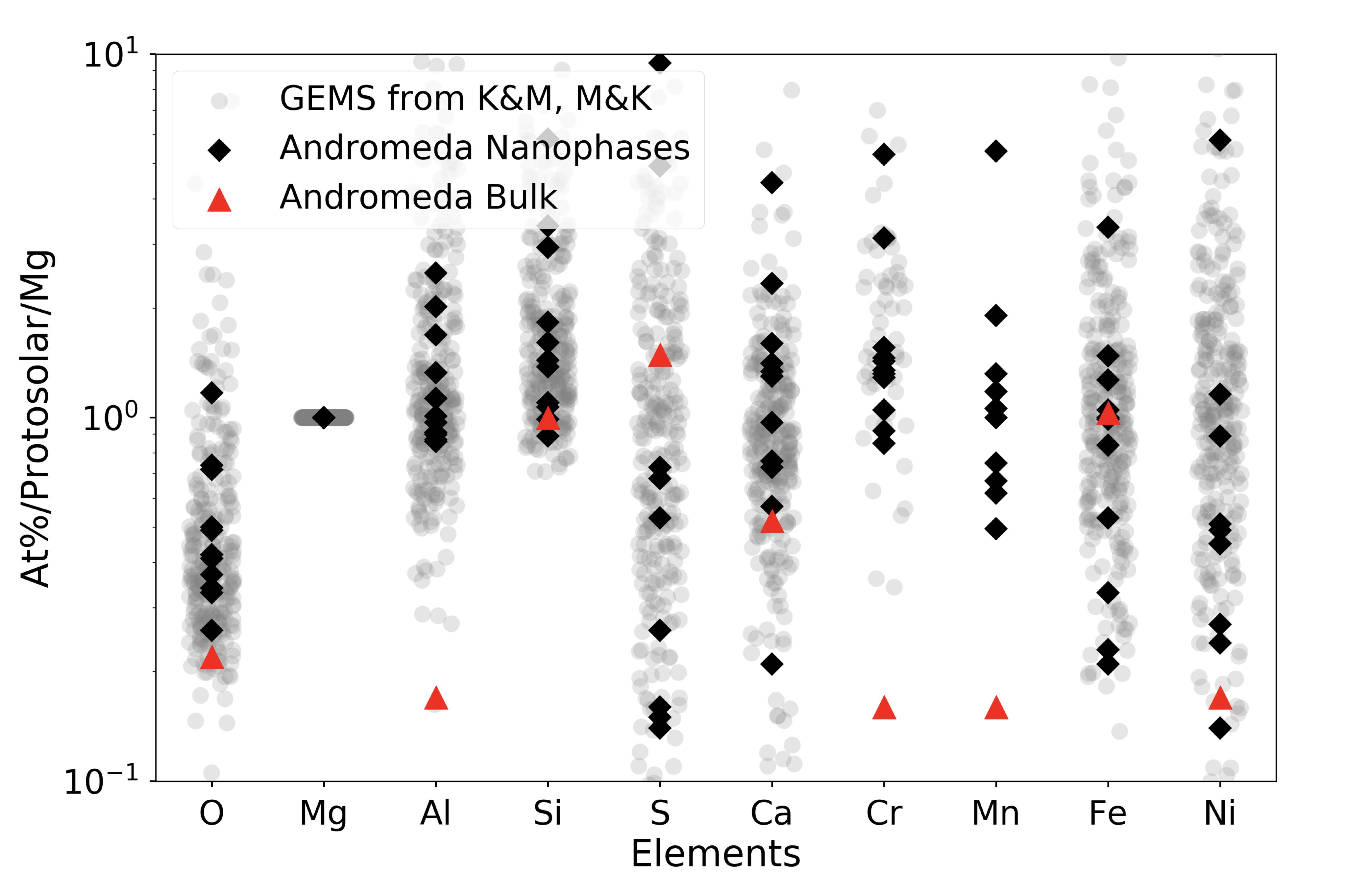}
  \caption{Figure \ref{fig:NanophaseCompositions}. Comparison of thermally modified and pristine GEMS-like and EA-like objects within Andromeda's fine grained material (FGM) to GEMS from literature.  Black diamonds indicate Andromeda nanophase object compositions normalized to protosolar and Mg.  Gray circles indicate GEMS compositions from \cite{Keller:2011bx} (K\&M) and \cite{Messenger:2015gz} (M\&K).  Red triangles show the bulk composition of the Andromeda FGM which is distinctly not GEMS-like since it contains pyroxenes, sulfides, and other astromaterials.}
  \label{fig:NanophaseCompositions}
\end{figure}

The compositions of GEMS measured by \citet{Keller:2011bx,Messenger:2015gz} are shown in Fig. \ref{fig:NanophaseCompositions} as gray points.
The Andromeda GEMS-like objects reported in this paper are also plotted in Fig. \ref{fig:NanophaseCompositions} and are consistent in elemental composition with GEMS.  
As expected, the bulk composition of the entire Andromeda FGM is not GEMS-like because it contains pyroxenes, sulfides, and other objects. 

\subsection*{Aerogel Capture: Dynamics and Thermal Environment}

\begin{figure}[t!]
  \centering
  \includegraphics[width=\ImageWidthFactor\textwidth]{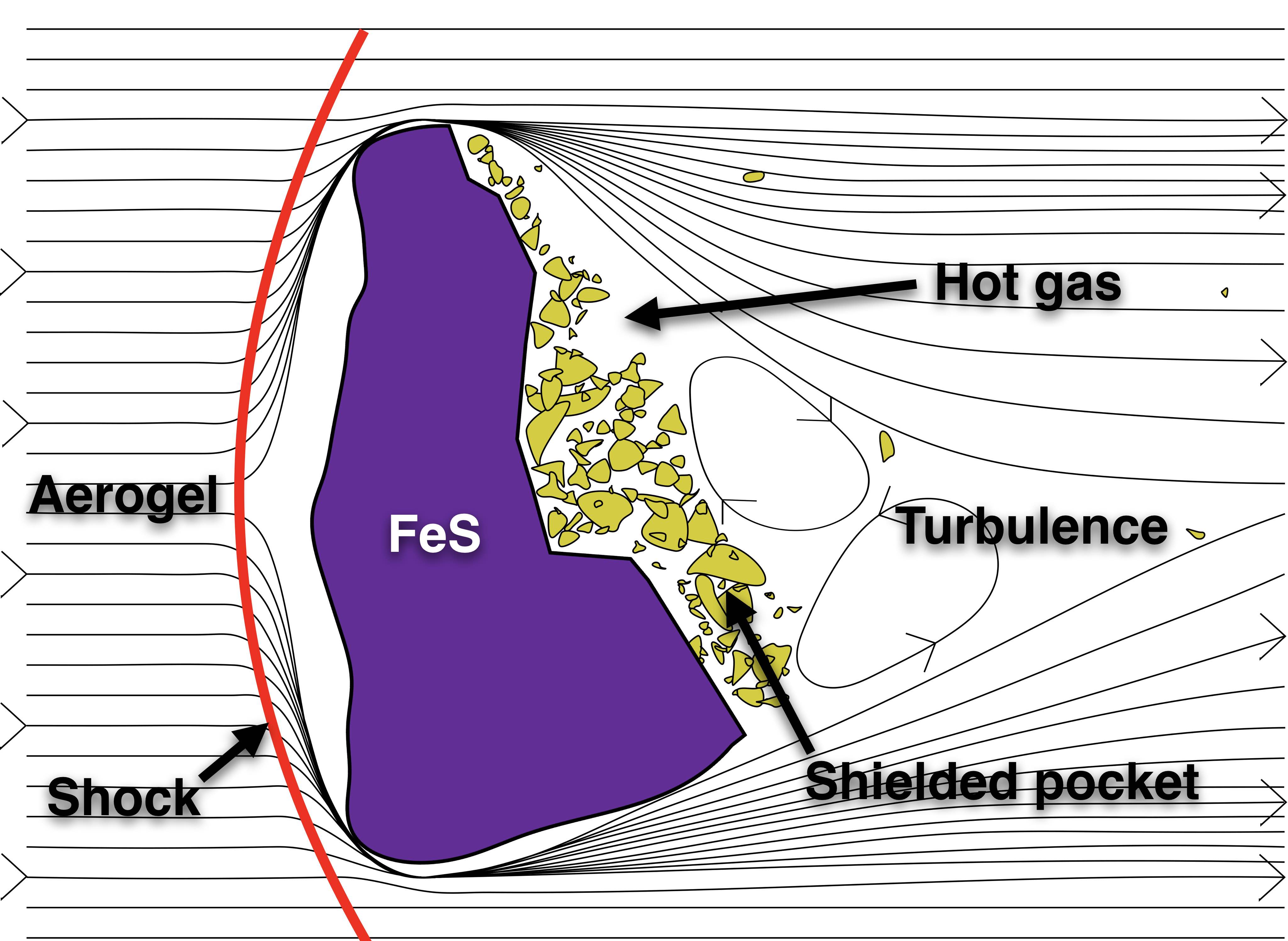}
  \caption{Figure \ref{fig:FlowCartoon}.  Hypersonic flow during capture.  Aerogel meets the face of Andromeda's sulfide producing a shock (red) that vaporizes aerogel and FeS.  Hot SiO$_2$ and FeS gas flow around the particle.  A turbulence region behind Andromeda causes some of the hot gas to enter the fine grained material (FGM) and heat it.  Some pockets of FGM are difficult to heat because they are shielded from the hot gas by other FGM.  Primitive objects in these pockets can survive capture. }
  \label{fig:FlowCartoon}
\end{figure}

Here we characterize the stopping dynamics and thermal environment experienced by Andromeda during capture in aerogel.
Following \cite{Dominguez:2009p21882}, we assume a hydrodynamic model for hypervelocity aerogel capture --- that is, the physics is identical to that of a projectile stopping in a gas (Fig. \ref{fig:FlowCartoon}).  
This approach is justified because the hydrodynamic forces are much larger than the mechanical strength of the aerogel.
In this regime, the range $R$ of a spherical particle with radius $r_g$, density $\rho$, and initial speed $v$, stopping in aerogel with density $\rho_a$, is 
 
\begin{equation}
	R = \lambda r_g \ln {v\over v_c}   
\end{equation}
where $v_c$ is the speed at which the mechanical strength of the aerogel becomes important, which we take
 to be comparable to the sound speed in aerogel, $\approx100$ m s$^{-1}$ \citep{Dominguez:2009p21882}, and $\lambda = (4/3) (\rho / \rho_a)$.
 
This calculation underestimates the actual range, because it only calculates the distance from the aerogel surface to the transition away from hydrodynamic stopping.  
After this transition, stopping is dominated by the crushing strength of the aerogel.   
At 6 km\,s$^{-1}$ capture speed, this residual range is short compared to the total range.

Because we have no information about the location of Andromeda within the original complex projectile --- not to mention the difficulty of modeling the stopping of a complex, fragmenting particle in aerogel ---  we treat Andromeda as an isolated, robust, refractory object. 
  
Andromeda was found at the end of the second-longest of at least 7 terminal tracks identified in track 191 (Fig. \ref{fig:TrackOptical}), a large track near one edge of Stardust cometary tile C2086 ($\rho_a = 0.028$ g\,cm$^{-3}$).
Since Andromeda is dominated by sulfide, we take  $\rho = 4.6$ g\,cm$^{-3}$,  so $\lambda \approx 220$.   With $v_i = 6.1$ km\,s$^{-1}$ and $r_g = 8\,\mu$m, the predicted range in this aerogel is 7.0\,mm.  
The observed range was 8.8\,mm, so the final $\approx1.8$\,mm of the range was probably dominated by non-hydrodynamic stopping, but may have included a range increment if Andromeda was not located near the leading edge of the track 191 projectile, so that it was delayed in encountering the aerogel stopping medium.   
 
We calculated that 70\% of the speed and 90\% of the kinetic energy were lost within the first 2\,mm of penetration into the aerogel, within $\approx$0.6\,$\mu$s after impact.
This is the region in which most of the heating occurs.
Because the projectile was hypersonic with Mach number $\gg10$, a strong shock just ahead of the leading edge of the projectile heated and vaporized the aerogel capture medium and the leading edge of the sulfide.   In the case of a strong shock ($M\gg1$), the post-shock temperature and pressure are
 \begin{equation}
  T \sim {2(\gamma-1) \over (\gamma+1)^2} {m \over k} v^2 = {5\over 36} {m \over k} v^2   
\end{equation}
and
\begin{equation}
  P \sim  {2 \over \gamma+1} \rho_a v^2 = {5\over 6} \rho_a v^2   
\end{equation}
where $m$ is the molecular mass,  $v$ is the velocity, $\rho_a$ is the aerogel density, $\gamma = c_p/c_v$ is the adiabatic index, $k$ is Boltzmann's constant and $\sim$ indicates an order of magnitude relationship \citep{Shu:1992vh}. Here we assume an adiabatic index $\gamma = 1.4$. This post-shock temperature corresponds to approximately 4 eV/molecule at maximum heating.

We assume a bulk aerogel temperature of $\sim200$\,K at the time of cometary dust capture. The impactor transfers energy that dissociates atoms and molecules from the aerogel and sulfide surface, and then heats the remaining solid and gas.
We computed silicate and FeS dissociation energies using Density Functional Theory (DFT) and from thermodynamic paramterizations in the JANAF database and show the results in Table \ref{tab:HypervelocityDissociation}.
If we further assume that the dissociation follows a Boltzmann distribution, i.e., $\exp^{-E_a/kT}$ where $kT$ = 4 eV, and $E_a$ is the dissociation energy computed by DFT, or derived from JANAF, then we expect one-quarter to one-third of the silica and FeS to dissociate into molecular SiO$_{2}$ and FeS.
We also expect a few percent each of O$_2$, O, Si, Fe and S.
This will lead to very high oxygen and sulfur fugacities.

\begin{figure}[hbt]
  \centering
  \includegraphics[width=\ImageWidthFactor\textwidth]{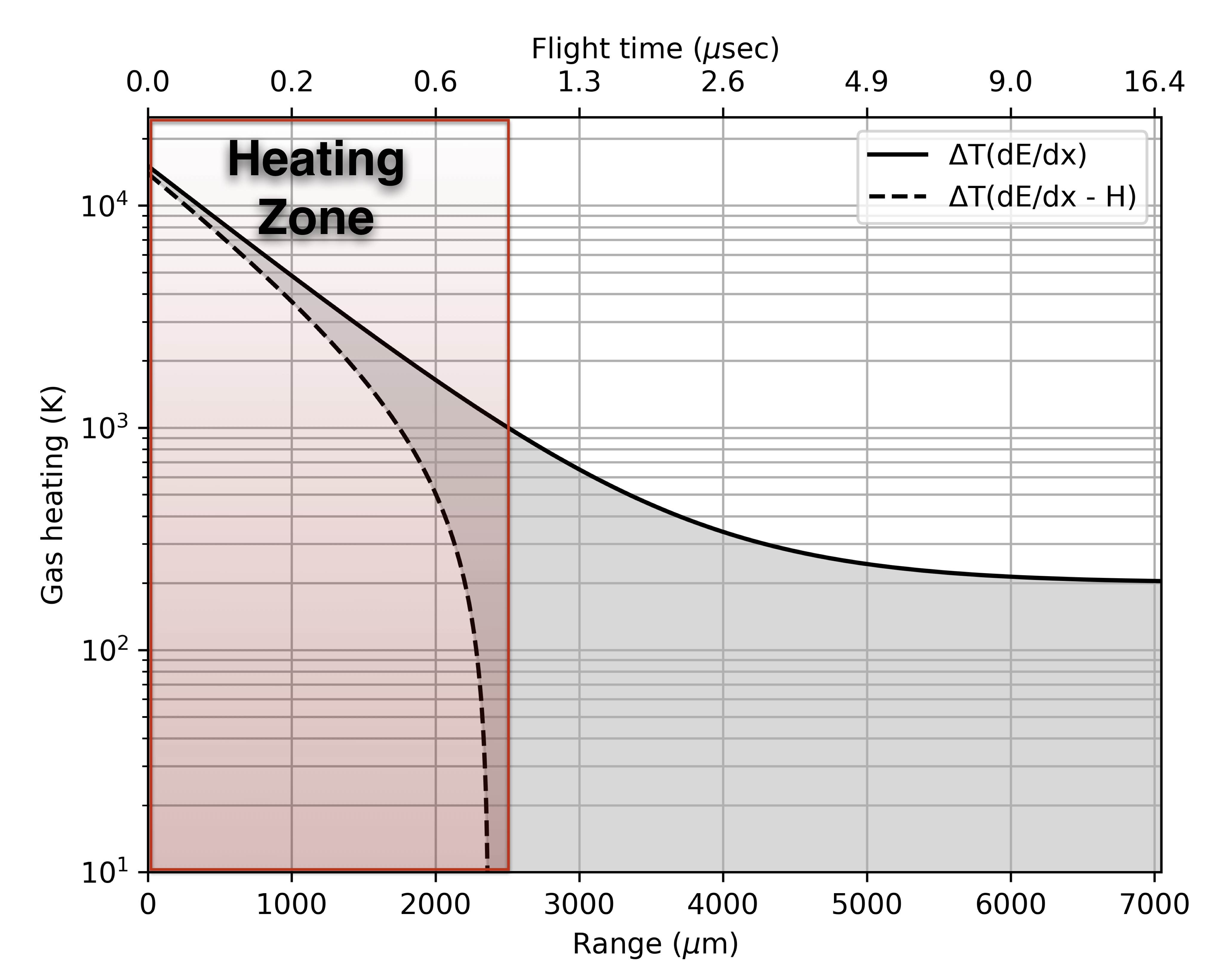}
  \caption{Figure \ref{fig:GasHeating}.  Vaporized gas temperature as a function of time.  The top solid line shows simple heating of gases and serves as an upper limit for the temperature.  The bottom dashed curve heats the gases only after accounting for the heat used to volatilize aerogel and sulfide, as well as to dissociate of FeS(g) $\rightarrow$ Fe(g) + S(g) and SiO$_2$(g) $\rightarrow$ SiO(g) + O(g) or Si(g) + O$_2$(g).  The bottom curve serves as a lower limit for the heating.  The shaded region shows the range of possible temperatures as a function of depth per the model.  The heating zone marks the region where gas temperatures exceed 1000\,K and lasts for < 1\,$\mu$s and < 3 mm into the track.}
  \label{fig:GasHeating}
\end{figure}

\begin{table*}[h]
\scriptsize
\caption{\label{tab:HypervelocityDissociation}}
\setlength\extrarowheight{1.7pt}
\centering{}
\begin{tabular}{lllll}
\multicolumn{5}{l}{{Table \ref{tab:HypervelocityDissociation}. Dissociation energies and abundances from DFT and JANAF at 0\,K.}}\\

\hline 
Reaction                                & \thead[l]{Energy \\(DFT, eV)} &\thead[l]{Dissociated \%\\(DFT)} & \thead[l]{Energy\\(JANAF, eV)} & \thead[l]{Dissociated \%\\(JANAF)}\\
\hline
SiO$_{2}$(s) $\rightarrow$ SiO$_{2}$(g)          & 6.5    & 20         & 6.2      & 21           \\
SiO$_{2}$(s) $\rightarrow$ SiO(g) + O(g)         & 12.1   & 5          & 10.9     & 7            \\
SiO$_{2}$(s) $\rightarrow$ Si(g) + O$_{2}$(g)    & 13.9   & 3          & 14.0     & 3            \\
SiO$_{2}$(s) $\rightarrow$ Si(g) + 2O(g)         & 22.5   & 0          & 19.1     & 1            \\
FeS(s) $\rightarrow$ FeS(g)                      & 5.1    & 28         & 4.9      & 29           \\
FeS(s) $\rightarrow$ Fe(g) + S(g)                & 12.2   & 5          & 8.2      & 13           \\

\multicolumn{5}{l}{$^{1}$Element for normalizing and subtracting the background.  See methods.}\\[2pt]

\end{tabular}
\end{table*}

We computed the temperature of the hot gas at the shockwave.  
We started with the energy loss due to deceleration.  
We assumed that the energy is partitioned equally between FeS and aerogel, so  $\approx$20\% of the silica and $\approx$30\% of the FeS  are volatalized, in accordance with the dissociation fractions predicted from DFT.  
The remaining solids and gases are heated to higher temperature by the remaining available energy.
The result is shown in Fig. \ref{fig:GasHeating}. The initial gas temperature is about 14000\,K at the start of the impact but falls rapidly.
The gas temperature falls to < 1000\,K by 2.5\,mm into the track after a time of < 1\,$\mu$s.
We define this initial region as the heating region (see Fig. \ref{fig:GasHeating}).
Although this model makes a number of simplifying assumptions,  the conclusion is clear:  this particle was exposed to $\approx10^4$\,K gas for $<1\,\mu$s.
 
The inertia of the radially-accelerated material and the pressure of the shocked gas were largely responsible for blowing the bulbous cavity in the aerogel (Fig. \ref{fig:TrackOptical}).
The travel time from the surface of the aerogel to the end of the hydrodynamic regime was $\approx$17\,$\mu$s (Fig. \ref{fig:GasHeating}).

 The heating of the sulfide and FGM in response to the $<1\,\mu$s  thermal pulse depends on their respective thermal inertias.  
 Given radius $r$, specific heat capacity $c_p$, density $\rho$, and thermal conductivity $k$, the thermal time constant $\tau$ is

\begin{equation}
  \tau \sim {\rho c_p \over k} r^2.
\end{equation}
 
Using physical constants for sulfide from \cite{Tsatis:1982tt}, we find that the thermal timescale for Andromeda's sulfide was $\tau > 20$\,$\mu$s.
The thermal time constant $\tau$ of a particle describes the time required for it to heat up or cool down to 1/$e \approx 36\%$ of the difference between its initial temperature and a new temperature.
This was an order of magnitude longer than the duration of its exposure to $10^4$\,K gas and limits the internal heating it would have experienced.
A 1\,$\mu$s exposure to $10^4$\,K gas with a time constant of 20\,$\mu$s leads to a maximum heating to $\approx$700\,K, which assumes no heating losses from ablation.  
Within 1\,$\mu$s after entering the aerogel, Andromeda was again in a cold environment, and frictional heating during stopping was negligible.
There were undoubtedly ablative losses as the sulfide decelerated, and these would have removed heat from the sulfide so the interior may have remained at a few hundred degrees for approximately its thermal time constant, $\approx$20\,$\mu$s.  
This surficial heating is reminiscent of entry of large meteorites by the Earth's atmosphere.

\subsection*{High-Ni bands in sulfide:  indicator of pre-accretional heating?}

The primary impactor was troilite with narrow Ni-rich bands of pyrrhotite 4C.  
In the case of the three bands averaged in Fig. \ref{fig:SulfideNiProfile}, we computed the width by assuming stoichiometry.   
First we assumed (Fe+Ni)/S = 1 outside the Ni-rich bands, and then calculated (Fe+Ni)/S = 0.994 at the center of the bands.
Pyrrhotite 4C has (Fe+Ni)/S = 0.875. Since the resolution of the EDS map was 30nm, 
the pyrrhotite bands should have been $\leq$ 2 nm or 1-2 unit cells thick (in the c* direction) and would have had a composition of Fe$_{6}$NiS$_{8}$ (Table \ref{tab:SulfideCompositions}, Sulfide\refspec{AndromedaIdealPyrrhotite}) to yield the observed composition.  Such a thin band would have produced a broadening of the diffraction reflections of $\approx$ 1 nm$^{-1}$.  
Based on SAED we found the peak broadening of the 4C superlattice reflections to be only 0.28 nm$^{-1}$ which could be produced by bands on the order of 6-7\,nm thick.
Some regions of the sulfide had Ni content as high as 1 at \% (i.e., double the Ni content in Fig. \ref{fig:SulfideNiProfile}) and the HRTEM image shown in Fig. \ref{fig:SulfideBands} shows a Ni band more than 20 nm thick.
The likely explanation is that the sulfide contained Ni-rich pyrrhotite bands of varying thickness averaging around 7 nm.

\cite{Gainsforth:2017er} related the broadening of pyrrhotite 4C superlattice reflections with varying degrees of disequilibrium.
If the sulfide is a nebular condensate (e.g. formed at 773\,K), then the disorder would record the initial formation conditions.
We follow the discussion of \cite{Gainsforth:2017er}, based on the experimental sulfidation of metal foils using H$_{2}$S by \cite{Lauretta:2005iea}.
Andromeda's [Ni/Fe]$_\odot$ = -1.3 implies a large depletion of Ni in the source material or a very low partition of Ni into the sulfide phase.
During sulfidation of Fe-metal by H$_{2}$S at 773\,K, Ni partitions primarily into the metal phase and forms taenite or can be pulled into schreibersite if P is present.
Lower formation temperatures lower the partitioning of Ni into the sulfide but formation out of equilibrium favors placing more Ni into the sulfide.
At 773\,K, for pyrrhotites showing a similar level of disorder, and for ``nebular metal'' with a few at \% Ni, Cr, P, and C, we expect a partition between metal and sulfide $D^{Ni/Fe}_{metal/FeS}$ = 1-2.
Therefore, this sulfide did not form under these conditions because its Ni content is one order of magnitude too low.
It could have formed under cooler temperatures but for a longer time so as to achieve a similar proximity to equilibrium.
At 673\,K, $D^{Ni/Fe}_{metal/FeS}$ = 5.
This would still put two to three times too much Ni into the sulfide.
Alternately, it could have formed from a Ni-poor  metal having only 0.1-0.5 \% Ni originally.

Another possibility is that Andromeda's sulfide is not a primary nebular condensate but rather formed in an igneous environment or experienced a metamorphic event.
The fact that the Ni has partitioned into well defined Ni-rich bands suggests that it was homogeneously dissolved throughout the sulfide at an earlier time, but later conditions pushed it out of equilibrium.
Such a condition could be obtained from a metamorphic event.

To understand the thermal history we need to know what thermal profile could have exsolved the pyrrhotite.
To our knowledge, the diffusion rate of Ni as a function of temperature in troilite has not been measured.  \cite{Condit:1974eo} measured the diffusion rates of Fe in troilite and pyrrhotite over the temperature range of 500 -- 1000\,K.    
We computed the timescales for diffusion of Ni by assuming it to have the same diffusivity as Fe.  
At 500\,K, the segregation might occur on the time scale of hours and at 1000\,K it could occur on the timescale of 1 second.
We did not extrapolate to temperatures below 413\,K because of a phase transition at that temperature.

We previously calculated that the sulfide was heated to no more than several hundred degrees for a few $\mu$s during aerogel capture, so diffusion is far too slow to have allowed the Ni-rich bands to form as a result of capture in aerogel.  We conclude that the bands are an original feature, either dating from initial formation or  are due to low-temperature metamorphism.

High-Ni lamellae from meteoritic sulfides with these characteristics have not previously been reported.
In previous analyses with similar sensitivity of four large Wild 2 sulfides, no such lamellae were observed.  This provides further evidence against an aerogel capture origin, and also points to a diversity in formation/alteration history of sulfides within Wild 2.
It is worth noting that the small variation in Ni content is very difficult to see instrumentally.
The lack of reports in the literature may be due to the fact that TEM instruments capable of seeing this phenomenon easily have only recently become available.
Our initial map of Andromeda missed the Ni-banding --- it was only after we acquired a high count map with > 10$^8$ counts, for Se quantification, that the bands became obvious.
As high-count EDS systems become more abundant in the community, more such banded sulfides may be found.

\subsection*{FGM Porosity, Density, Modal abundance}

Here we compare the physical characteristics of the FGM to FGM in CP-IDPs and to {\em in situ} cometary observations. 

We estimated the porosity ($\phi$ = fraction of filled space) of the FGM from Si and S EDS maps. We could not determine the modal abundance of carbonaceous material because of the ubiquitous epoxy.   
We found $\phi=0.57$ for the FGM.  
88\% of the material in the FGM was silicate and 12\% of the material was sulfide, so we estimated the bulk density of the FGM to be $\approx$1.4 g/cm$^3$.  

\cite{Gainsforth:2017ts} found porosities and densities for several IDPs and found values in approximate agreement with Andromeda.
Specifically, the porosity, density and silicate/sulfide abundance of the FGM is comparable to the silicate/sulfide abundance in the "Nessie" IDP (L2071 Cluster 17).
\cite{Joswiak:2007ve} also reported densities of 0.7-1.7 g/cm$^{3}$ for several TEM sections of IDPs, and \cite{Fraundorf:1982vd} reported densities of 0.7-2.2 g/cm$^3$ for IDPs measured on a quartz fiber balance.
As noted in \cite{Gainsforth:2017ts}, measurements from the Rosetta mission using the GIADA instrument found that dust grains ejected from Comet 67P/Churyumov-Gerasimenko (C-G) had densities between 1-3 g/cm$^3$\citep{Rotundi:2015ur}. 

Using high-precision Doppler measurements, the Rosetta mission was able to measure the mass of 67P/Churyumov-Gerasimenko nucleus \citep{Sierks:2015uo}.  This yields a density of 0.5 g/cm$^{3}$.
The discrepancy between this bulk measurement of the comet nucleus and the higher densities seen in CP-IDPs and C-G particles implies the existence of void spaces on a larger length scale than the size of CP-IDPs or cometary particles collected by Stardust.

\subsection*{FGM Heating}
 
As discussed earlier, vaporized aerogel and sulfide gas would have been in thermal contact with the leading edge of the particle.  
Unless the particle was rotating very rapidly ($\approx10^6$ rotations s$^{-1}$),  Andromeda did not rotate significantly during the 1\,$\mu$s hot thermal pulse, so one side was a leading edge and the opposite side a trailing edge.  
The FGM in the sulfide embayment was likely on the trailing edge of the sulfide during capture in aerogel. 
The hydrodynamic pressure (> 1 GPa on the leading edge) was probably much larger than the binding strength between the FGM particles, so it is unlikely that this loosely-bound material would have survived intact if it had been exposed directly to the flow of hot, vaporized gas.

In the results section, we reported evidence of variable heating of individual silicate particles in the FGM.
Here we explore two possible heating mechanisms, related to aerogel capture.

The most likely mechanism was direct thermal contact with gas from the shocked, vaporized mixture of aerogel and ablated upstream cometary materials.  
Cooler gas in the cavitation region of the supersonic flow could have penetrated the porous FGM and been in thermal contact with it for the duration of the $<1$\,$\mu$s hot pulse (Fig. \ref{fig:FlowCartoon}).  
Near the top of the track, thej SiO$_2$ gas would have been partially dissociated into SiO and O, and the FeS gas would have been partially dissociated into Fe and S, so it would have been oxidizing and sulfidizing. 
The thermal time constant of $\approx$300\,nm silicates is small, $\ll0.1$\,$\mu$s, so any particles exposed to the gas would have rapidly equilibrated with it, but variations in permeability of the porous FGM could be expected to lead to variable exposures to the hot gas.
We would expect some regions of the FGM to be inaccessible to the gas and any GEMS, EAs or other primary objects in these ``shielded pockets" would be unheated or minimally heated, and remain unoxidized and unsulfidized.
Without more information about the permeability and porosity of the FGM, it is probably not possible to model this process with sufficient fidelity to compare with detailed observations.   

We also considered radiative heat transfer from a hot, ionized gas, conservatively assuming that the radiating gas was an optically thick blackbody, and the FGM was also a blackbody with 100\% absorptivity.  
We calculated that the heating of FGM was less than 1\,K and rule it out as a significant heating mechanism.

Excess silica is a common indicator of aerogel mixing in the bulb region of Stardust tracks \citep{2008M&PS...43...97L,Stodolna:2012ht}.  The bulk SiO$_{2}$ component of the FGM is only 1 at\% in excess of the SiO$_{2}$ content of olivine -- a relatively SiO$_{2}$ poor mineral.
This means that the amount of SiO$_{2}$ contamination of the FGM is probably small -- on the order of a percent.
Since some gas phase SiO$_{2}$ would have been produced during aerogel capture, this significantly limits the heating dose within the FGM and explains why some primitive components survived.
We computed the heating that would be experienced by 200\,K quartz (as a proxy for FGM) from thermal equilibration with an SiO$_{2}$ gas at $10^4$\,K.  
Assuming the gas/solid ratio is 1\%, we found that the FGM should heat to $\approx$300\,K.
Apparently, some FGM heated more than that.
One possibility is that more gas was present and heated the sample, but later diffused out without condensing.
If the abundance of gas was an order of magnitude greater, then the FGM could have heated to about 1100\,K which is high enough to begin to see alteration -- though only the smallest objects (e.g, GEMS) would alter on the timescale of 1\,$\mu$s.
A more likely explanation is that the hot gas heated a smaller fraction of the material because it was only in contact with the pores.
Indeed, if we assume that the gas was only in contact with about 10\% of the FGM, then it should have heated about 10\% of the FGM to 1100\,K and then deposited a total mass of 1\% Si when it condensed.

\cite{Brownlee:2005wh} found that heating GEMS to temperatures $\geq$ 1000\,K for several hours results in sub-solidus crystallization of the silicate and they postulated that equilibrated aggregates could have formed by heating in the nebular environment.
If this picture is correct, then equilibrated aggregates should be more robust to transient heating events than GEMS -- i.e., GEMS should be more susceptible to destruction during aerogel capture than equilibrated aggregates. 
With this in mind, the survival of a GEMS would provide the strongest constraint on the temperatures during heating.  

\subsection*{Primitive FGM}

FGM\refspec{GEMS1} is the best candidate for a GEMS in Andromeda (Fig. \ref{fig:GEMS}).  It contains euhedral sulfides, does not contain excess silica, and retains the expected chemical composition of a GEMS.
While there is likely some heating on one side of FGM\refspec{GEMS1} where there is one subhedral external sulfide, and sulfur has bled into the silicate matrix, the appearance of FGM\refspec{GEMS1} is similar to GEMS seen in IDPs, many of which also show evidence of heating from atmospheric entry or nebular events \citep{1993LPI....24..205B, 2005ASPC..341..668D}.   
FGM\refspec{GEMS1} is indistinguishable in its characteristics with GEMS in CP-IDPs.  
For example, see Fig. 5 in \cite{Keller:2005vo} or Fig. 1 in \cite{Keller:2011bx}.

FGM\refspec{EA1} and FGM\refspec{EA2} appear to be indistingishable from equilibrated aggregates in CP-IDPs, with no indicators of shock heating from the aerogel capture.
Equilibrated aggregates are probably more robust to heating than GEMS and so it is possible that they were heated but not modified by the process.
For comparison to EAs in IDPs, see Fig. 2 in \cite{Rietmeijer:2009vm}, Fig. 2 in \cite{Keller:2005vo} or Fig. 1 in \cite{Keller:2009wa}.

\subsection*{Altered FGM}

We argue that the remaining nanophases studied in Andromeda's FGM were thermally altered to a greater or lesser degree on account of the presence of one or more of the heating indicators we described in the previous sections.

FGM\refspec{Slag3} was especially noteworthy because EDS imaging could clearly distinguish euhedral grains of distinct compositions but brightfield imaging and diffraction indicated the grains were amorphous.
We interpret this as similar to "shadow grains" described previously by \cite{Stodolna:2012ht} and \cite{2008M&PS...43...97L} when a crystalline phase is amorphized by rapid heating and cooling but not sufficiently heated for the atoms to diffuse away.
The atoms then settle into metastable amorphous phases with shapes and compositions very close to their original crystalline counterparts.

FGM\refspec{Slag3} was in contact with the primary impactor sulfide in a region containing many Fe-oxides, and did not show excess silica.  
Because it was found in a central part of the FGM, it would have been unlikely that it came into contact with hot, molten aerogel.  
Instead its morphology was consistent with the hot gas model we propose.
If hot gas permeated the FGM, then it could have penetrated all open pores and come in contact with the sulfide where it would have rapidly cooled.

\subsection*{Amorphous FeO$_x$}

The amorphous Fe-oxides found near FGM\refspec{Alt4} (Fig. \ref{fig:Alt4}), FGM\refspec{Alt5} and FGM\refspec{Slag3} may be important to understanding the origin of the FGM.  
The oxide is embedded in epoxy so it is not possible to determine the exact oxygen abundance even after subtracting off a background epoxy spectrum.  
However, based on rough stoichiometry we expect a minimum oxidation of Fe$^{2+}$, and in some cases more oxygen is present than would be expected for Fe$^{3+}$ so even oxy-hydroxide is consistent with our error margin.
Other rock forming elements are present in lower abundance.  

The Fe-oxide is closest in composition to magnetite rims in some mildly heated IDPs.
In such IDPs it is possible to find magnetite rims on the order of 100 nm which contain nanocrystalline and amorphous oxidized Fe.
These often contain small abundances of other elements deriving primarily from the neighboring sulfides and silicates from which they formed.
The oxides in Andromeda all contain significant abundances of other elements, but Si and S are present in every spectrum and furthermore are present in roughly equal amounts of a few At\%.

Experiments by \cite{JosephANuth:SmokeSynthesis} and \cite{Rietmeijer:wi} have shown that it is possible to synthesize Fe oxide amorphous nanocrystals directly from vapor phase in an $\approx$ 90 Torr H$_{2}$-rich gas (0.12 bar).  
They were formed at 500 - 1500\,K and on time scales "much less than a second."
With the exception of the H$_{2}$ atmosphere and the low pressures, these conditions are consistent with what we should expect from shock heating during aerogel capture.

Fig. \ref{fig:FeOxideAnalogs} shows nanocrystalline magnetite in an IDP alongside artificial smokes produced by \cite{JosephANuth:SmokeSynthesis}.  The inset shows an Fe-oxide from Andromeda for comparison (also cf. Fig. \ref{fig:Alt4}).

\cite{Barth:2017dq} show that at lower temperatures, e.g. 450 $^{\circ}$C, Fe-oxides can form from a high $f_{O_2}$ and high $f_{S_2}$ gas.
Specifically under high fugacity, low temperature conditions, oxides are stable unless log($f_{O_2}$) < log($f_{S_2}$) by about 20 log units. 
Our calculations above show that log($f_{O_2}$) $\approx$ log($f_{S_2}$) which means that hematite would have been the stable phase if aerogel + sulfide gas were condensing.
Because the process would have been so rapid, the Fe-oxide would not have crystallized but instead solidified as an amorphous solid.

The prevalence of Fe-oxide at the interface between the large sulfide impactor and the FGM would have been expected.
Because the sulfide did not have time to heat up more than a few hundred degrees during the capture process, it should have acted as a heat sink for nearby material.
As hot gas permated through the FGM and came into contact with the sulfide, it would have cooled to a few hundred degrees which would have driven rapid condensation.
Hot gas further out in the FGM would have diffused out of the FGM at a later point and would likely leave less residue.

\subsection*{Wild 2 as an Aggregate Rock}

Immediately after the Stardust return, \cite{Brownlee:2006kw} proposed that comet Wild 2 is, effectively, an aggregate of inner and outer Solar System materials which indicated substantial mixing throughout the solar nebula.
The amount of presolar material present in Wild 2 is consistent with CP-IDPs, and other primitive bodies \citep{Floss:2013hd}, so Wild 2 is certainly sampling some unaltered solar nebula material.
Work on large rocky objects (terminal particles) in Wild 2 has frequently identified chondrules, and refractory rocks indicating that inner Solar System materials composed a significant fraction of the captured mass of Wild 2 \citep{Joswiak2012:thetome, 2008Sci...321.1664N, Brownlee:2013gv, Gainsforth:2015jv}.
Recently, \cite{Joswiak:2017dk} found that CAI-like fragments account for $\approx$ 1 vol\% in cometary samples including Wild 2 and a giant cluster IDP named U2-20.
\cite{Joswiak:2017dk} make a strong case that U2-20 has a cometary origin, and have noted (personal communication) that it has coarse grained rocks (CAI-like) in direct contact with FGM including GEMS and EAs.
\cite{Wooden:2017eV} recently tied together several lines of evidence pointing to late formation of Wild 2 materials, several Mya the onset of CAI formation.
These include an abundance of objects with O isotopic signatures close to the terrestrial value, and an overabundance of late stage objects including Fe-rich type II chondrules, and late $^{26}$Al dates. 
Wild 2 olivines also show some evidence of having been metamorphosed. \citep{Frank:2014efa}. 

Nevertheless, while Wild 2 FGM has occasionally been found near large terminal particles, the FGM has usually been heavily altered and as a consequence, comparison with FGM in CP-IDPs has been difficult.
For example, the only enstatite whisker found to date in Wild 2 is encased in glass -- a phonomenon never reported previously \citep{Stodolna:2013co}.
Febo has had to date the most promising FGM, but significant unresolved questions about the alteration of the FGM remain \citep{Joswiak2012:thetome}.  
A long-standing open question has therefore been whether Wild 2 has any ``CP-IDP" FGM at all.
The fortuitous survival of the Andromeda FGM indicates that the answer is ``yes''.

The presence of sulfides in the FGM with a significantly different composition than the impactor indicate that they are not simply ablation products from the impactor during capture.
This means that the FGM does not share the same formation/alteration history that Andromeda's large sulfide experienced.
If the sulfide had been metamorphosed while aggregated to the FGM, the FGM would not have survived, so either the sulfide formed originally with Ni-rich bands, or aggregated with the FGM after metamorphosis/nebular heating.
This observation supports the conclusion that Wild 2 is an aggregate rock with GEMS and equilibrated aggregates alongside larger rocks/minerals.

K and P are volatile elements and are present in nearly all the objects within the FGM material that we studied, and may provide another window into the relationship between the FGM and the coarse material.
Because of their volatility, the gas heating model would predict that K and P would disperse throughout the FGM with the gas and then recondense as the gas cooled.  However, K is significantly enriched in the bulk composition with [K/Si]$_\odot$ = 0.32.
Simple redistribution of K would predict that the bulk would remain unchanged or even depleted as K escaped into the nearby aerogel.
High K has been noted previously within Stardust samples \citep{Flynn:2006p4738}, though there is some evidence for K as an occasional contaminant in aerogel alongside Cl and Ca \citep{Rietmeijer:2015dm}.
K is especially intriguing as it is commonly enriched in more evolved materials.
The carrier phase for the K in Andromeda has not been identified and may have been destroyed during capture.
This suggests that an investigation of the abundance and carrier of K in CP-IDPs may yield interesting results.

P, on the other hand, is not particularly enriched in the bulk FGM ([P/Si]$_\odot$ = 0.049 $\approx$ 0), but is significantly overabundant around the nanophases we investigated (typical [P/Si]$_\odot$ = 0.15 with the highest value at FGM\refspec{Alt5} with [P/Si]$_\odot$)=0.55).
The nanophases we investigated were preferentially those which looked like GEMS or EAs or other primitive material (i.e. not large crystals, sulfides, etc.)
Therefore, while FGM\ref{Slag4} shows that P was likely redistributed to some degree during capture, it must also be more concentrated within the GEMS/EA-type objects or else it would have also been more concentrated in the bulk.
The reason for this is not clear but warrants investigation in the future.

\section*{CONCLUSIONS}

We summarize key conclusions:
\begin{enumerate}
	\item Wild 2 contains objects that are indistinguishable from GEMS and equilibrated aggregates found in CP-IDPs.
	\item Wild 2 is an aggregate containing both large crystals and FGM.
	\item Ni-banding in the sulfide impactor may show evidence for heating between formation and  incorporation into comet Wild 2.
	\item The distribution of K and P leads us to suggest that future research would benefit by identifying the origin and chemistry of K- and P- bearing phases in IDPs.
\end{enumerate}

We have also constrained the processes active during Andromeda's capture, which allows us to better resolve which processes can be attributed to cometary/Solar System processes and which are artifacts of capture.
We demonstrate that objects within the fine grained material (FGM) behind the sulfide impactor were heated by a transient hot gas on the order of $\approx 10^4$\,K but for 1\,$\mu$s at most.
This is sufficient to melt individual silicates but some objects appear to have survived in a relatively pristine condition.  This is likely due to variations in the porosity and the thermal inertia of nearby objects (e.g. the impactor and large crystalline grains) which shielded some FGM from exposure to the hot gas. 
The FGM only has excess aerogel in select regions. 
We have identified the gas-phase condensation of Fe-oxide from volatilized Fe and O produced during aerogel capture.

\section*{ACKNOWLEDGMENTS}

Work at the Molecular Foundry and Advanced Light Source was supported by the Office of Science, Office of Basic Energy Sciences, of the U.S. Department of Energy under Contract No. DE-AC02-05CH11231.
This work was supported under the LARS program by NASA grant NNX16AK14G.
MZ was supported by NASA's Emerging Worlds Program.

We wish to thank the staff at Johnson Space Center for making IDPs and Stardust samples available to the community.
We wish to thank Karen Bustillo and the rest of the staff at the National Center for Electron Microscopy for providing spectacular instrumentation without which this paper would not exist.  
We wish to thank Yufeng Liang and David Prendergast at the Molecular Foundry for tutelage with DFT and access to computing resources.
We wish to thank Natasha Johnson and Joseph A. Nuth III for supplying the Fe smokes used for comparison with  Fe-oxides.

\renewcommand\refname{REFERENCES}

\bibliographystyle{MAPS}
\addcontentsline{toc}{section}{\refname}\bibliography{PapersLibraryTrimmed}

\onecolumn
\begin{landscape}


\begin{table*}[h]
\scriptsize
\caption{\label{tab:NanophaseCompositionsProtosolar}}
\setlength\extrarowheight{1.7pt}
\centering{}
\begin{tabular}{llllllllllllllllll}
\multicolumn{18}{l}{{Table \ref{tab:NanophaseCompositionsProtosolar}. Nanophase compositions of capture modified objects from TEM EDS Normalized to Mg and CI.}}\\

\hline 

& \multicolumn{14}{c}{Atomic \% / Protosolar / (Mg or Si)$^1$} & &\\
 
\cline{2-15}

\# &     O &    Na &          Mg &    Al &          Si &      P &     S &      K &    Ca &    Ti &    Cr &     Mn &    Fe &    Ni &  $\tilde{\chi}^2_{7}\enskip^{(2)}$ &  $p^{(3)}$ \\
\hline
 FGM\refspec{FGMBulk} &  0.22 &  0.27 &  0.08 &  0.17 &  {\em 1.00} &  1.12 &  1.49 &  2.08 &  0.52 &  b.d. &  0.16 &  0.16 &  1.03 &  0.17 &  2.44 &  0.98 \\
 FGM\refspec{GEMS1} &  0.41 &  0.20 &  {\em 1.00} &  1.33 &  1.61 &  1.06 &  0.73 &  4.06 &  0.73 &  1.24 &  1.05 &  0.62 &  0.53 &  0.45 &  0.55 &  0.20 \\
 FGM\refspec{EA1} &  0.34 &  0.53 &  {\em 1.00} &  1.01 &  1.10 &  1.30 &  0.06 &  3.83 &  0.57 &  0.79 &  0.92 &  1.18 &  0.33 &  0.24 &  1.31 &  0.76 \\
 FGM\refspec{EA2} &  0.26 &  0.34 &  {\em 1.00} &  0.86 &  0.99 &  1.41 &  0.06 &  2.70 &  0.21 &  0.73 &  5.30 &  1.00 &  0.23 &  0.04 &  17.78 &  1.00 \\
 FGM\refspec{Alt1} &  0.74 &  0.18 &  {\em 1.00} &  2.02 &  3.37 &  b.d. &  9.45 &  5.84 &  1.41 &  1.59 &  1.46 &  5.41 &  1.27 &  5.80 &  3.02 &  1.00 \\
 FGM\refspec{Alt2} &  0.33 &  0.96 &  {\em 1.00} &  0.97 &  0.89 &  1.818 &  0.14 &  0.97 &  1.34 &  0.29 &  1.35 &  0.495 &  0.21 &  0.27 &  1.42 &  0.81 \\
 FGM\refspec{Alt3} &  0.37 &  b.d. &  {\em 1.00} &  0.90 &  1.07 &  1.16 &  0.15 &  1.45 &  0.76 &  1.40 &  1.32 &  0.75 &  0.84 &  0.49 &  0.84 &  0.45 \\
 FGM\refspec{Alt4} &  0.49 &  0.74 &  {\em 1.00} &  1.69 &  1.83 &  1.59 &  0.68 &  2.88 &  2.34 &  0.71 &  1.43 &  1.32 &  1.05 &  0.51 &  0.92 &  0.51 \\
 FGM\refspec{Alt5} &  0.42 &  0.31 &  {\em 1.00} &  0.87 &  1.38 &  4.95 &  0.16 &  3.17 &  1.30 &  0.46 &  3.12 &  1.06 &  1.00 &  0.05 &  3.12 &  1.00 \\
 FGM\refspec{Slag1} &  0.72 &  0.76 &  {\em 1.00} &  2.50 &  2.94 &  1.43 &  0.53 &  5.92 &  4.43 &  2.15 &  1.56 &  0.67 &  1.48 &  1.16 &  1.95 &  0.94 \\
 FGM\refspec{Slag2} &  1.17 &  0.69 &  {\em 1.00} &  0.91 &  5.84 &  1.318 &  4.92 &  2.54 &  0.97 &  b.d. &  0.85 &  1.91 &  3.34 &  0.89 &  1.58 &  0.87 \\
 FGM\refspec{Slag3} &  0.50 &  0.77 &  {\em 1.00} &  1.13 &  1.44 &  3.14 &  0.26 &  5.94 &  1.60 &  1.07 &  1.29 &  b.d. &  0.99 &  0.14 &  1.02 &  0.58 \\
 FGM\refspec{Slag4} &  1.64 &  0.50 &  {\em 1.00} &  7.50 &  8.49 &  14.06 &  1.17 &  39.82 &  2.36 &  8.76 &  9.73 &  b.d. &  5.60 &  5.01 &  2.52 &  0.99 \\

\hline 

\multicolumn{18}{l}{$^1$ The FGM bulk is the only spectrum that has been normalized to Si.  The FGM as a whole is subchondritic in Mg, but approximately chondritic in Si.  See text.}\\
\multicolumn{18}{l}{$^2$ $\tilde{\chi}^2_{7}$ is the reduced $\chi^2$ with 7 degrees of freedom, showing the difference between this spectrum and an ideal GEMS composition.  See text.}\\
\multicolumn{18}{l}{$^3$ $p$ is the probability that this measurement is inconsistent with the distribution of GEMS compositions.}\\

\end{tabular}
\end{table*}


\begin{table*}[t]
\begin{centering}
\caption{\label{tab:PhoneNumbers}}

\setlength\extrarowheight{1.7pt}
\newcommand{\tablewidthphone}{5}

\par\end{centering}

\centering{}
\scriptsize
\begin{tabular}{lllll}
\multicolumn{5}{l}{Table \ref{tab:PhoneNumbers}. Sample ID numbers.}\\[2pt]

\hline 

Table & Spectrum & UC Berkeley ID & $\Sigma$ Peak Counts \\[2pt]
 
\hline 

\ref{tab:SulfideCompositions} & Sulfide\newspec{AndromedaSulfideThinBlade} & 20160229 Andromeda G1,S1 thin blade & $6.1\cdot10^4$ & \\
\ref{tab:SulfideCompositions} & Sulfide\newspec{AndromedaSulfideBulk}      & 20151203 Andromeda G1,S1 - Stack 9 - Full Area Corrected & $3.2\cdot10^8$ \\
\ref{tab:SulfideCompositions} & Sulfide\newspec{AndromedaSulfideFeMatrix}  & 20151203 Andromeda G1,S1 - Stack 9 - Full Area Corrected & $3.1\cdot10^6$ \\
\ref{tab:SulfideCompositions} & Sulfide\newspec{AndromedaSulfideNiBand}    & 20151203 Andromeda G1,S1 - Stack 9 - Full Area Corrected & $9.1\cdot10^5$ \\
\ref{tab:SulfideCompositions} & Sulfide\newspec{AndromedaSulfideNiBand2}   & 20151203 Andromeda G1,S1 - Stack 9 - Full Area Corrected & $4.4\cdot10^5$ \\
\ref{tab:SulfideCompositions} & Sulfide\newspec{AndromedaIdealPyrrhotite}  & Calculated in discussion                                 & N/A \\
\ref{tab:SulfideCompositions} & Sulfide\newspec{CPXSulfide}                & 20151203 Andromeda G1,S1 - Stack 5 - Sulfide & $1.1\cdot10^4$ \\
\ref{tab:SulfideCompositions} & Sulfide\newspec{FeboSulfide}   & 20131219 - Febo & $6.7\cdot10^7$ \\
\ref{tab:NanophaseCompositions} &  FGM\newspec{FGMBulk} &  20151203 Andromeda - Stack 2 - FGM Bulk &  3.8$\cdot10^6$ \\
 \ref{tab:NanophaseCompositions} &  FGM\newspec{GEMS1} &  20160229 Andromeda -Mg-C &  1.4$\cdot10^6$ \\
 \ref{tab:NanophaseCompositions} &  FGM\newspec{EA1} &  20151203 Andromeda - Stack 3 - EA1  &  1.8$\cdot10^6$ \\
 \ref{tab:NanophaseCompositions} &  FGM\newspec{EA2} &  20160229 Andromeda -Mg-D &  7.3$\cdot10^5$ \\
 \ref{tab:NanophaseCompositions} &  FGM\newspec{Alt1} &  20160229 Andromeda -Mg-B &  1.2$\cdot10^6$ \\
 \ref{tab:NanophaseCompositions} &  FGM\newspec{Alt2} &  20151203 Andromeda - Stack 4 - Daisy &  1.2$\cdot10^5$ \\
 \ref{tab:NanophaseCompositions} &  FGM\newspec{Alt3} &  20160130 Andromeda - Duck2 sum stacks &  8.8$\cdot10^5$ \\
 \ref{tab:NanophaseCompositions} &  FGM\newspec{Alt4} &  20160130 Andromeda - Duck1 Stack 3 &  5$\cdot10^5$ \\
 \ref{tab:NanophaseCompositions} &  FGM\newspec{Alt5} &  20160229 Andromeda -Mg-E &  3.9$\cdot10^5$ \\
 \ref{tab:NanophaseCompositions} &  FGM\newspec{Slag1} &  20160229 Andromeda -Mg-F &  5.6$\cdot10^5$ \\
 \ref{tab:NanophaseCompositions} &  FGM\newspec{Slag2} &  20160229 Andromeda -Mg-A &  1.6$\cdot10^6$ \\
 \ref{tab:NanophaseCompositions} &  FGM\newspec{Slag3} &  20151203 Andromeda - Stack 6 - Robodog &  1.3$\cdot10^6$ \\
 \ref{tab:NanophaseCompositions} &  FGM\newspec{Slag4} &  20151203 Andromeda - Stack 7 &  4.9$\cdot10^5$ \\
\ref{tab:FeOxides} &  FeOx\newspec{Alt4-FeOxide1} &  20160130 Andromeda - Duck1 Stack 3 &  1.9$\cdot10^5$ \\
 \ref{tab:FeOxides} &  FeOx\newspec{Alt4-FeOxide2} &  20160130 Andromeda - Duck1 Stack 3 &  3.9$\cdot10^4$ \\
 \ref{tab:FeOxides} &  FeOx\newspec{Alt4-FeOxide3} &  20160130 Andromeda - Duck1 Stack 3 &  1$\cdot10^4$ \\
 \ref{tab:FeOxides} &  FeOx\newspec{Alt4-FeOxide4} &  20160130 Andromeda - Duck1 Stack 3 &  1.7$\cdot10^4$ \\
 \ref{tab:FeOxides} &  FeOx\newspec{Alt4-FeOxide5} &  20160130 Andromeda - Duck1 Stack 3 &  1.3$\cdot10^4$ \\
 \ref{tab:FeOxides} &  FeOx\newspec{Alt5-FeOxide6} &  20160229 Andromeda - Mg-E - Fe Core &  1.1$\cdot10^4$ \\
 \ref{tab:FeOxides} &  FeOx\newspec{Alt5-FeOxide7} &  20160229 Andromeda - Mg-E - Fe Side &  7$\cdot10^3$ \\
 \ref{tab:FeOxides} &  FeOx\newspec{Lambda-Magnetite} &  20170809 - Lambda B4 S7 - Stack 6 - Magnetite &  3.6$\cdot10^5$ \\
\ref{tab:PyroxeneCompositions} & Px\newspec{AndromedaCPX} & 20151203 Andromeda G1,S1 - Stack 5 - CPX& $9.9\cdot10^5$ \\
\hline 

\multicolumn{5}{l}{$^{a}$The thickness correction has been chosen to optimize for troilite.}\\[2pt]

\end{tabular}
\end{table*}

\end{landscape}


\end{document}